\documentclass[twocolumn,aps,pre]{revtex4}
\usepackage{makeidx}
\usepackage{amsmath}
\usepackage{amssymb}
\usepackage{graphicx}
\usepackage{times}

\begin{document}

\title{\sffamily\bfseries\large Statics and dynamics of elastic manifolds in media
with long-range correlated disorder}
\author{ \sffamily\bfseries\normalsize  Andrei A. Fedorenko, Pierre Le Doussal,
 and Kay J\"{o}rg Wiese \smallskip\smallskip}

\affiliation{CNRS-Laboratoire de Physique Th{\'e}orique de l'Ecole Normale Sup{\'e}rieure,%
 24 rue Lhomond, 75231 Paris,  France}
\date{January 20, 2007}

\begin{abstract}
We study the statics and dynamics of an elastic manifold in a
disordered medium with quenched defects correlated as $\sim r^{-a}
$ for large separation $r$. We derive the functional
renormalization-group equations to one-loop order, which allow us to
describe the universal properties of the system in equilibrium and
at the depinning transition. Using a double $\varepsilon=4-d$ and
$\delta=4-a$ expansion, we compute the fixed points characterizing
different universality classes and analyze their regions of
stability. The long-range disorder-correlator remains analytic but
generates short-range disorder whose correlator exhibits the usual
cusp. The critical exponents and universal amplitudes are computed
to first order in $\varepsilon$ and $\delta$ at the fixed points.
At depinning, a velocity-versus-force exponent $\beta$ larger than
unity can occur. We discuss possible realizations using extended
defects.
\end{abstract}

\maketitle

\section{Introduction}
\label{sec1}

Elastic objects in random media are the simplest example of a
disordered system exhibiting metastability, glassy behavior, and
dimensional reduction, which are difficulties present in a broader
class of disordered systems
\cite{fisher-phys-rep98,kardar-phys-rep98,brazovskii03}. They can be
used to model a remarkable set of experimental systems.  Domain
walls in magnets behave as elastic interfaces and can experience
either random bond disorder (RB) as in ferromagnets with
nonmagnetic impurities, or random field disorder (RF) as in
disordered antiferromagnets in an external magnetic field
\cite{domain-walls-exp}.  The interface between two immiscible
liquids in a porous medium exhibits the same behavior and undergoes
a depinning transition as the pressure difference is increased
\cite{fluids-invasion}. Charge-density waves (CDW) in solids show a
similar conduction threshold \cite{cdw}.  Another example of
periodic systems are vortex lines in superconductors which can form
different glass phases in the presence of weak disorder
\cite{vortex,nattermann-00,bragg}.  In all these systems, the
interplay between elastic forces that tend to keep the system
ordered, i.e., flat or periodic, and quenched disorder, which
promotes deformations of the local structure, forms a complicated
energy landscape with numerous metastable states.  This results in
glassy properties and a nontrivial response of the system to
external perturbations.  In particular, the interface becomes rough
with displacements growing with the distance $x$ as
\begin{equation}
  C(x)\sim x^{2\zeta},\label{def-roughness}
\end{equation}
where $\zeta$ is the roughness exponent. Elastic periodic structures
in the presence of disorder lose their strict translational order
and form quasi-long-range order characterized by a slow growth of
displacements,
\begin{equation}
  C(x) = \mathcal{A}_d \ln |x| \label{def-roughness-periodic},
\end{equation}
where the amplitude $\mathcal{A}_d$ is universal in the simplest
case. At zero temperature, a driving force $f$ exceeding the
threshold value $f_c$ is required to set the elastic manifold into
steady motion with a velocity $v$ that vanishes as $v\sim
(f-f_c)^{\beta}$ at the transition point. The correlation length
diverges close to the transition $f=f_c$ as $\xi\sim (f-f_c)^{-\nu}$
and the characteristic time as $\tau \sim \xi^z$, where $z$ is the
dynamic critical exponent. Note that the roughness exponent and the
universal amplitudes determined at the depinning transition are in
general different from the exponent and amplitudes measured in
equilibrium.

Two methods were developed to study the statics of an elastic manifold
in a disordered medium. One of them is the Gaussian variational
approximation (GVA) performed in replica space, which can be applied to both
classes of elastic manifolds, i.e., to interfaces \cite{mezard90} without overhangs
and to periodic systems \cite{bragg,korshunov93}.  Within this approach,
which is believed to be exact in the mean-field limit, i.e., when the
manifold lives in a space of infinite dimensions, metastability is
described by breaking of replica symmetry, which allows one to compute
the static correlation functions and to obtain different thermodynamic
properties. Another method that can be applied to dynamics as well as
to statics is the functional renormalization group (FRG)
\cite{fisher86}.  Simple scaling arguments show that large-scale
properties of elastic systems are governed by disorder for
$d<d_{\mathrm{uc}}=4$ and that perturbation theory in the disorder
breaks down on scales larger than the so-called Larkin scale
\cite{larkin70}. To overcome this difficulty, one performs a
renormalization-group analysis. It was shown that in this case one has
to renormalize the whole disorder correlator that becomes a
nonanalytic function beyond the Larkin scale
\cite{fisher86,nstl92,lnst07,narayan-fisher93}. The appearance of a
non-analyticity in the form of a cusp at the origin is related to
metastability, and nicely accounts for the generation of a threshold
force at the depinning transition.  It was recently shown that the FRG
can unambiguously be extended to higher loop order so that the
underlining nonanalytic field theory is probably renormalizable to
all orders \cite{chauve01,ledoussal02,ledoussal04}. Although the two
methods, GVA and FRG, are very different, they provide a fairly consistent picture
of the statics, and recently a relation between them was found
\cite{ledoussal03-3}. There is also good agreement with results of
numerical simulations, not only for critical exponents
\cite{roters-pre2002-1,roters-pre2002-2,rosso2003} but also for the
whole renormalized disorder correlator \cite{middleton06}. However,
many questions remain open. Although the dynamics in the vicinity of
the depinning transition and at zero temperature is well understood,
there is no satisfactory theory for finite temperature, and in
particular for the thermal rounding of the depinning transition
\cite{chauve-00}. It is also remarkable that the exponent $\beta$ in
experiments on depinning is usually larger than $1$, while FRG and
numerical simulations of elastic systems with weak disorder give
values smaller than~$1$.

Most studies of elastic manifolds in a disordered medium treat
uncorrelated pointlike disorder. Real systems, however, often contain
extended defects in the form of linear dislocations, planar grain
boundaries, three-dimensional cavities, etc. It is known that such
extended defects, or pointlike defects with sufficiently long-range
correlations, can change the bulk critical behavior
\cite{dorogovtsev-80,boyanovsky-82,lawrie-84,fedorenko-04,%
weinrib-83,korucheva-98,fedorenko-00}. Flux lines
in superconductors are the most prominent example. The pinning of the
flux lines by disorder prevents the dissipation of energy and
determines the critical current $J_c$, which is of great importance
for applications. It was found that extended defects produced, for
instance, by heavy-ion irradiation, can increase $J_c$ by several
orders of magnitude \cite{civale-91}.  Systems with anisotropic
orientation of extended defects can be described by a model in which
all defects are strongly correlated in $\varepsilon_d$ dimensions and
randomly distributed over the remaining $d-\varepsilon_d$
dimensions. The case $\varepsilon_d=0$ is associated with uncorrelated
pointlike defects, while extended columnar or planar defects are
related to the cases $\varepsilon_d=1$ and 2, respectively. The
bulk-critical behavior in the presence of this type of disorder was
studied in Refs.~\cite{dorogovtsev-80,boyanovsky-82,lawrie-84,fedorenko-04}
using a perturbative RG analysis
in conjunction with a double expansion in $\varepsilon=4-d$ and
$\varepsilon_d$. The pinning of flux lines by columnar disorder was
studied in Ref.~\cite{nelson-92}, where it was shown that the system
forms a Bose glass phase with flux lines strongly localized on the
columnar defects, resulting in a zero dc linear resistivity. It was
argued recently that the topologically ordered glass phase (Bragg
glass) formed by flux lines can be destroyed in the vicinity of a
single planar defect \cite{emig06}. It has been shown that the small
dispersion in orientation of columnar defects forms a new
thermodynamic phase called ``splayed glass'' \cite{hwa-93}. In this
phase, the entanglement of flux lines enhances significantly the
transport of superconductors \cite{civale}. Competition between
various types of disorder, point and columnar, has also been
studied, at equilibrium \cite{hwa2,chauvethesis} and in the moving
phases \cite{chauvembg}.

In the case of an isotropic distribution of disorder, power-law
correlations are the simplest example with the possibility for a
scaling behavior with new fixed points (FPs) and new critical
exponents. The bulk-critical behavior of systems in which defects
are correlated according to a power law $r^{-a}$ for large
separation $r$ was studied in
Refs.~\cite{weinrib-83,korucheva-98,fedorenko-00}. The power-law
correlation of defects in  $d$-dimensional space with exponent
$a=d-\varepsilon_d$  can be ascribed to randomly distributed
extended defects of internal dimension $\varepsilon_d$ with random
orientation. For example, $a=d$ corresponds to uncorrelated
pointlike defects,  $a=d-1$ ($a=d-2$) describes infinite lines
(planes) of defects with random orientation. In general, one would
probably not expect a pure power-law decay of correlations. However,
if the correlations of defects arise from different sources with a
broad distribution of characteristic length scales, one can expect
that the resulting correlations will over several decades be
approximated by an effective power law~\cite{weinrib-83}. If the
correlation function of disorder can be expressed as a finite sum of
power-law contributions $\sum_i c_i r^{-a_i}$, one can expect that
the scaling behavior is dominated by the term with the smallest
$a_i$~\cite{weinrib-83}. Power-law correlations with a noninteger
value $a=d-d_f$ can be found in systems containing defects with
fractal dimension $d_f$ \cite{yamazaki-88}. For example, the
behavior of $^4\mathrm{He}$ in aerogels is argued to be described by
an \textit{XY} model with LR correlated defects \cite{vascuez-03}. This is
closely related to the behavior of nematic liquid crystals enclosed
in a single pore of aerosil gel, which was  recently studied in
Ref.~\cite{feldman-04}, using the approximation in which the pore
hull is considered a disconnected fractal. Finally, studies of the
Kardar-Parisi-Zhang (KPZ) equation with power-law correlations in
time \cite{kpzcorr} bear connections to the case $d=1$ considered
here. However, the perturbative method used there cannot address
directly the zero-temperature (strong KPZ coupling) phase,
contrary to our present study.

In the present paper, we study the statics and dynamics of elastic
manifolds in the presence of (power-law) LR correlated disorder using
the FRG approach to one-loop order. The paper is organized as
follows. Section \ref{sec2} introduces the model. Possible
physical realizations are considered in Sec. \ref{sec3}.  Section
\ref{sec4} describes the dynamical formalism and perturbation theory.
In Sec. \ref{sec5}, we renormalize the theory and derive the FRG
equations to one-loop order.  In Sec. \ref{sec6}, we study random
bond, in Sec. \ref{sec7}, random field, and in Sec. \ref{sec8},
periodic disorder. In Sec. \ref{sec:extended} we discuss fully
isotropic extended defects. In the final section, we summarize the
obtained results and our conclusions.

\section{The model}
\label{sec2}

We consider a $d$-dimensional elastic manifold embedded in a
$D$-dimensional space  with quenched disorder. The configuration
of the manifold is described by an $N$-component  displacement
field denoted below $u(x)$, or equivalently $u_x$, where $x$
denotes the $d$-dimensional internal coordinate of the manifold.
For example, a domain wall corresponds to $d=D-1$ and $N=1$, a CDW
to $d=D$ and $N=1$, and a flux lattice to $d=D$ and $N=2$. In what
follows, we focus for simplicity on the case $N=1$ and elastic
objects with short-range elasticity. Extensions to $N>1$ and LR
elasticity are straightforward for the statics. The energy of the
manifold in the presence of disorder is defined by the Hamiltonian
\begin{equation}\label{Hamiltonian}
  \mathcal{H} = \int d^d x \left[ \frac{c}2 [\nabla u(x)]^2 + V\big(x,u(x)\big)  \right ],
\end{equation}
where $c$ is the elasticity and $V$ is a random potential. In this
paper, we study the model where the second cumulant of the random
potential has the form
\begin{eqnarray}
  \overline{V(x,u)V(x',u')}&=&R_1(u-u')\delta^d(x-x')  \nonumber \\
 &&+ R_2(u-u')g(x-x'). \label{model}
\end{eqnarray}
The first part corresponds to pointlike disorder with short-range
(SR) correlations in internal space. The second part corresponds
to long-range (LR) disorder in internal space and the function $g(x)
\sim x^{-a}$ at large $x$ with $a>0$. For convenience, we normalize it so that
its Fourier transform is $g(q)=|q|^{a-d}$ at small $q$ with unit
amplitude.  \textit{A priori} we are interested in the case $a<d$, where the
correlations decay sufficiently slowly in internal space. We denote
everywhere below $\int_q=\int \frac{d^dq}{(2\pi)^d}$ and
$\int_x=\int d^dx$. The short-scale uv cutoff is implied at $q \sim
\Lambda$ and the size of the system is $L$.

One could start with model (\ref{model}), setting $R_1=0$;
however, as we show below, a nonzero $R_1$ is generated under coarse
graining. Note that the functions $R_i(u)$ can themselves \textit{a priori}
be SR, LR, or periodic in the direction of the displacement field
$u$. For SR disorder in internal space only, i.e., $R_2=0$, these
cases are usually referred to as random bond (RB), random field
[$R_1(u) \sim |u|$ at large $u$] (RF), and random periodic (RP)
universality classes. Below we discuss how these classes extend
to the case of LR internal disorder ($R_2$ nonzero).

The model (\ref{Hamiltonian}) and (\ref{model}) could easily be
studied using presently available numerical algorithms for directed
manifolds, in its statics (e.g., exact ground-state determinations) and
its dynamics (e.g., critical configuration at depinning), by directly
implementing a random potential correlated as described by Eq.
(\ref{model}). It is also interesting to examine which type of
correlations in a random medium can naturally lead to Eq. (\ref{model})
and how such disorder could be realized from, e.g., distributions of
extended defects, since some of them may be experimentally feasible.

\section{Realizations and universality classes}
\label{sec3}

\subsection{Defect potential}

Let us first recall how long-range correlations can arise in the
potential created by defects. To this purpose, call $v(r)$ the defect
potential, in the simplest case taken to be proportional to
defect density. Consider for simplicity a large number of weak
defect lines with a uniform and isotropic distribution in a space of
dimension $D$. These create an almost Gaussian random potential
$v(r)$ with
\begin{eqnarray}
\overline{v(r) v(r')} \sim
\frac{v_{\mathrm{LR}}^2}{|r-r'|^a} \qquad\mbox{for }{r \to \infty} \label{disorder potential}
\end{eqnarray}
and $a=D-1$. To derive this, consider defects of finite radius $a_d$.
The probability that point $r'$ is contained in the defect going
through $r$ is $\sim (a_d/|r-r'|)^{D-1}$, i.e., inversely
proportional to the sphere of radius $|r-r'|$. This is easily
generalized to isotropic distributions of extended defects of
internal dimension $\varepsilon_d$, with $a=D-\varepsilon_d$.
Note that by
extended defects we mean defects that are perfectly correlated
along their internal dimension. Generalizations where defects are
themselves (anisotropic) fractals can also be considered.

An important case is a uniform  distribution of extended defects
in $D$-dimensional space, but isotropic only within a
linear subspace of dimension $D'$. For instance, one can irradiate a
material in the bulk while simultaneously rotating it along an axis.
This produces a distribution of linear defects ($\varepsilon_d=1$),
isotropic within the plane ($D'=2$), and normal to the axis
(see Fig.\ \ref{fig-dw-inplane}). More generally, this yields a defect
potential with second cumulant,
\begin{eqnarray}
&& \overline{v(r,z) v(r',z)} = g(r-r') f(z-z'), \label{splitted}   \\
&& g(r) \sim r^{-a} \nonumber\ ,
\end{eqnarray}
while $f(z)$ is short-ranged (here $r \in R^{D'}$, $z \in R^{D-D'}$,
$a=D'-\varepsilon_d$).

Although we mostly discuss extended defects, other sources of long-range
correlations are possible, such as defects where each single one creates a
long-ranged disorder potential, or a substrate matrix itself
quenched at a critical point.

\subsection{Coupling to the manifold}
We now examine how the long-range correlated defect potential
couples to the elastic manifold and what type of LR model results. A
general formulation of this coupling (see, e.g., \cite{brazovskii03})
has the form
\begin{equation}
V(x,u)=\int d^{D-d} z\, v(x,z) \rho (x,z,u), \label{potential}
\end{equation}
where the defect potential lives in the $D$-dimensional space
parametrized by $(x,u)$ and $x \in R^d$ is the internal coordinate
of the manifold. $\rho(x,z,u)$ is the manifold density.
Each type of coupling to the disorder corresponds
to a different function $\rho(x,z,u)$, and we now indicate the main
cases.

\subsubsection {Elastic interfaces in random bond disorder }

\begin{figure}[tbp]
\includegraphics[clip,width=2.2 in]{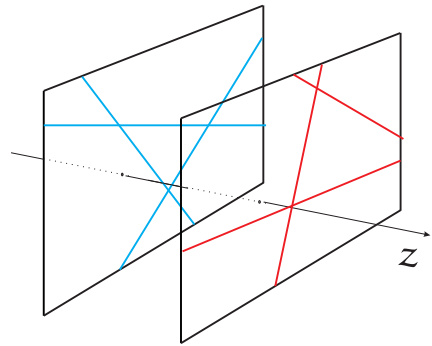}
\caption{(Color online) Linear defects randomly and isotropically
distributed on parallel planes
with random distances between them. This geometry mimics distribution
(\ref{splitted}).}
\label{fig-dw-inplane}
\end{figure}

Let us first discuss elastic interfaces in the so-called random
bond (RB) case, where the coupling between disorder and interface occurs
only in the vicinity of the interface as, e.g., for domain walls in
magnets with random bond disorder. This corresponds to the
choice
\begin{equation}
  \rho(x,z,u) \sim \delta(z-u), \label{RB-potential}
\end{equation}
hence the additional variable $z$ introduced in Eq. (\ref{potential}) is
identical to $u$, the displacement field (with in general $D-d=N$ ).
In that case,
\begin{equation}
V_{\mathrm{RB}}(x,u) \sim v(x,u).  \label{rb}
\end{equation}

Consider now a uniform distribution of defects in the
$D$-dimensional plane but {\it isotropically distributed within the
(averaged direction) of the internal space of the manifold}. This is given
by Eq. (\ref{splitted}) above with $D'=d$,
\begin{eqnarray}
\overline{V_{\mathrm{RB}}(x,u) V_{\mathrm{RB}}(0,0)} = g(x) R_2(u),
\label{bare-correlator}
\end{eqnarray}
which is model (\ref{model}) with a SR function $R_2(u)$ and, in
full generality, $a=d-\varepsilon_d$. The physical realization in terms
of extended defects is thus an interface ($d=2$) in $D=3$ with line
defects all orthogonal to the $u$ directions, isotropically
distributed within the (average) plane of the interface, and $a=1$.
This is illustrated in Fig. \ref{fig-dw-inplane}.

Another physical realization consists of extended defects with
finite random lengths such that the distribution of lengths has a
power-law tail for large lengths. For instance, needles of variable
lengths aligned along one direction could act on a directed polymer
$d=1$ as power-law correlated disorder in internal space.

An interesting, though qualitatively different, case occurs when the
extended defects are distributed isotropically in the whole $(x,u)$
space. This will be discussed in Sec. \ref{sec:extended}. Finally,
note that we consider weak Gaussian disorder. It is
possible that at strong disorder another phase exists where the line
or manifold gets localized along the strongest extended defect.

\subsubsection {Elastic interfaces in random field disorder }

Random field (RF) disorder is described by the function
\begin{equation}
   \rho(x,z,u)\sim \Theta(u-z), \label{RF-potential}
\end{equation}
where $\Theta(z)$ is the Heaviside step function. This means that
the change in energy when the interface moves between two
configurations is proportional to the sum of all defect potentials in
the volume (in $R^D$) spanned by this change. The discussion of the
geometry of defects needed to produce LR disorder in internal space
is identical to the last section. Substitution  of
Eq.~(\ref{RF-potential}) into Eq.~(\ref{potential}) yields the RF
disorder correlator, which can be approximated by Eq.~(\ref{model})
with $R_i(u) \sim - u$ for large~$u$.

\subsubsection {Periodic systems}

As an example of  periodic systems, we consider incommensurate
single-\textit{Q} CDWs. In that case $D=d$, hence the function
$\rho(x,z,u)=\rho(x,u)$ in Eq. (\ref{potential}). The electron density
of CDWs neglecting effects caused by an applied strain has the
form~\cite{cdw,brazovskii03}
\begin{equation}
   \rho(x,\phi)\sim\rho_0+\rho_1\cos\{2k_{\mathrm{F}} [x_{\perp}-u(x)]\},
    \label{CDW-potential}
\end{equation}
where the displacement $u(x)$ of the maximum of the density is
related to the standard phase field via
$\phi(x)=-2k_{\mathrm{F}}u(x)$, where $k_{\mathrm{F}}$ is the Fermi
wave vector. The $d$-dimensional space is split into
$x=(x_{\parallel}, x_{\perp})$,  with $x_{\perp}$ denoting the
modulation direction of the CDW and $k_{\mathrm{F}}$  the Fermi
wave vector.

We again consider the situation of extended defects all aligned with
the direction $x_{\parallel}$ and isotropically distributed in that
subspace. The random potential experienced by the CDW is given by
\begin{eqnarray}
  V(x,\phi)=h_1(x)\cos\phi(x)+h_2(x)\sin\phi(x), \label{V-CDW}
\end{eqnarray}
with Gaussian distributed $h_1(x)=v(x)\cos(2k_{\mathrm{F}}
x_{\perp})$ and $h_2(x)=v(x)\sin(2k_{\mathrm{F}} x_{\perp})$. On
large scales $k_{\mathrm{F}} x_{\perp} \gg 1$, and their cumulant can be
approximated by [from Eq. (\ref{splitted})]
\begin{eqnarray}
 \overline{ h_i(x) h_j(0)} = \frac12 v^2_{\mathrm{SR}}\delta_{ij} \delta^d(x)
 + \frac12 \frac{v_{\mathrm{LR}}^2}{x_{\parallel}^a} \delta_{ij} \delta(x_{\perp}),
 \label{h-cumul}
\end{eqnarray}
where we have  omitted all rapidly fluctuating contributions.
Equations~(\ref{V-CDW}) and (\ref{h-cumul}) give the potential correlator in a form
that can be generalized to
\begin{eqnarray}
&& \!\!\!\!\!\!\!\!
 \overline{V(x,u)V(x',u')} = R_1(u-u')\delta^d(x-x')  \nonumber \\
 && \ \ \ \ \ \ \ \ \ \ \ \ \ \ \ \ \ \ \ \ \ \
 +  R_2(u-u')g(x_{\parallel}-x_{\parallel}')\delta^{d_{\perp}}(x_{\perp}-x_{\perp}'),\qquad
 \label{cdw-cumulant}
\end{eqnarray}
with $d_{\perp}=1$ and bare functions $R_i(\phi)=\frac12 v_i^2 \cos
(\phi)$, $u\equiv\phi$. Thus periodic systems are described by
periodic functions $R_i(u)$. Here $d_{\perp}$ is the dimension of
the transverse subspace. Note that the Hamiltonian
$\mathcal{H}_{\mathrm{\textit{XY}}}= \int d^d x [\frac12(\nabla \phi)^2 +V
(x,\phi)]$ with $V (x,\phi)$ given by Eq.~(\ref{V-CDW}) and a Gaussian
distribution of fields $\overline{h_i(x)h_j(x')}\sim g(x-x')$
describes the \textit{XY} model with long-range correlated random fields.
Therefore, the latter can be mapped onto periodic manifolds with
correlator  (\ref{cdw-cumulant}) and $d_{\perp}=0$, i.e., to model
(\ref{model}) with periodic functions $R_i(u)$. In the next section,
we will show how the FRG picture of model (\ref{cdw-cumulant})
can be obtained from the FRG results for model (\ref{model}). It is
worthwhile to note that in the case of periodic systems, the integration
in Fourier space is supposed to be over the first Brillouin zone.
Note also that we have neglected the coupling of disorder to the
long wavelength part of the density $-\rho_0\int d^d x v(x) \nabla u(x)$
as it is usually irrelevant near the upper critical dimension.
Indeed, in the replicated Hamiltonian (see below), this coupling generates
additionally to the SR term
$-1/T\int d^d x \sigma_1 \nabla u_a(x)\nabla u_b(x)$
the LR term
\[
-\frac1{T} \int d^d x\, d^dx' \sigma_2 g(x_{\parallel}-x'_{\parallel})
\delta^{d_{\perp}}(x_{\perp}-x_{\perp}') \nabla u_a(x) \nabla u_b(x').
\]
For small disorder in the vicinity of the upper critical dimension,
both of them renormalize to zero according to
\begin{eqnarray}
d_{\ell}\sigma_1 &=& (2-d-2\zeta)\sigma_1+..., \\
d_{\ell}\sigma_2 &=& (2-a-d_{\perp}-2\zeta)\sigma_2+...
\end{eqnarray}

\section{Dynamical formalism}
\label{sec4}

The overdamped dynamics of the elastic manifold in a disordered medium can be described
by the equation of motion
\begin{equation} \label{eq-motion}
\eta\partial_t u_{x t} = c\nabla^2 u_{x t} + F(x,u_{x t})+f_{xt},
\end{equation}
where $\eta$ is the friction coefficient. In the presence of an  applied
force $f$, the center-of-mass velocity is $v=L^{-d}\int_x\partial_t u_{xt}$. The
pinning force reads $F=-\partial_u V(x,u) $, and thus, for correlator
(\ref{model}), the second cumulant of the force
is given by
\begin{eqnarray}
\overline{F(x,u)F(x',u')}&=&{\Delta}_1(u-u')\delta^d(x-x')  \nonumber \\
 &&+ {\Delta}_2(u-u')g(x-x'),
\end{eqnarray}
with $\Delta_i=-R''_i(u)$ in the bare model. In the following, we will always
use $g(q)= |q|^{a-d}$ and $g(x)=\int_{q} e^{iqx} g(q)$.

The most important quantity of interest is the roughness exponent $\zeta$ measured
in equilibrium or at the depinning transition $f=f_c$ defined by
\begin{equation}\label{roughness}
  C(x-x')=\overline{|u_x-u_{x'}|^2} \sim |x-x'|^{2\zeta}.
\end{equation}
The velocity vanishes at the depinning transition as $v\sim |f-f_c |^{\beta}$,
while the correlation length diverges at the transition as
$\xi \sim |f-f_c |^{-\nu}$. One can also introduce the dynamic critical exponent $z$,
which relates spatial and temporal correlations via $t \sim x^z$.

Let us briefly sketch how one can construct the perturbation theory
in disorder. We adopt the dynamic formalism. It also allows us to
obtain the statics equations (to one loop and $N=1$ these can easily
be deduced, as can be checked using replica). Instead of a direct
solution of the equation of motion (\ref{eq-motion}) with consequent
averaging over different initial conditions and disorder
configurations, we employ the formalism of generating functional.
Introducing the response field $\hat{u}_{xt}$ we derive the
effective action, which reads
\begin{eqnarray}
\!\!\!\!\!\! S &=&\int_{x t} i\hat{u}_{x t}(\eta\partial_t - c\nabla^2 +m^2
)u_{x t}-
\int_{x t} i\hat{u}_{x t} f_{x t} \nonumber \\
&& - \frac12\int_{x t t'}\, i\hat{u}_{x t}i\hat{u}_{x t'}
\Delta_1({u}_{x t}-{u}_{x t'})  \nonumber \\
&& - \frac12\int_{xx' t t'}\, i\hat{u}_{x t}i\hat{u}_{x' t'}
g(x-x') \Delta_2({u}_{x t}-{u}_{x' t'}),\ \  \label{action}
\end{eqnarray}
where we have added a small mass $m$, which plays the role of an IR
cutoff.  To study the critical domain, one has to take the limit $m \to
0$.  The average of the observable $A[u_{xt}]$ over dynamic
trajectories with different initial conditions and over different
disorder configurations can be written as follows:
\begin{equation}
\langle A[u_{xt}] \rangle = \int \mathcal{D}[u] \mathcal{D}[\hat{u}] A[u_{xt}]
e^{-S[u,\hat{u}]}.
\end{equation}
Furthermore, the response to the external perturbation $f_{xt}$, which is
local in time and in space,  can be computed using
$\langle A[u_{xt}] i \hat{u}_{xt} \rangle = \frac{\delta}{\delta f_{xt}} \langle A[u_{xt}]  \rangle  $.
Note that causality is fulfilled,  and here we adopt the Ito convention, which
results in getting rid of all closed loops composed of response functions.

\begin{figure}[tbp]
\includegraphics[clip,width=2.2 in]{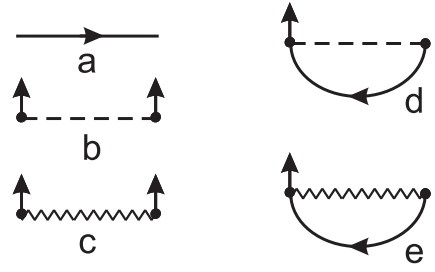}
\caption{Diagrammatic rules: $a$, propogator; $b$, SR disorder vertex; $c$,
LR disorder vertex; $d$ and $e$,  one-loop diagrams generating the critical force
at the depinning and giving correction to the  mobility and elasticity.}
\label{fig-mobility}
\end{figure}

In the absence of  LR correlated disorder action, Eq.
(\ref{action}) exhibits the so-called statistical tilt symmetry
(STS), i.e., the invariance of the disorder terms under the tilt
$u_{xt} \to u_{xt} + h_x$ with an arbitrary function $h_x$. The STS
gives the exact identity $\int_t \mathcal{R}_{qt}=1/cq^2$ for the
response function $\mathcal{R}_{qt}=\langle u_{qt}\ i\hat{u}_{-q0}
\rangle$, which implies that the elasticity is uncorrected by
disorder to all orders. LR correlated disorder destroys the STS of
action (\ref{action}), and thus, in principle, allows for a
renormalization of the elasticity. The quadratic part of the action
(\ref{action}) yields the free response function
\begin{eqnarray} \label{propagator}
 \langle u_{qt}\ i\hat{u}_{-q0}  \rangle_0 = R_{qt}=\frac{\Theta(t)}{\eta}
  e^{-(cq^2+m^2)t/\eta},
\end{eqnarray}
which can be used to generate the perturbation theory in disorder.
The theory has two disorder interaction vertices $\Delta_1(u)$ and
$\Delta_2(u)$. At each vertex $\Delta_i(u)$ there is one
conservation rule for momentum and two for frequency while each
vertex $\Delta_2(u)$ carries additional momentum dependence. In what
follows, we generalize  the splitted diagrammatic method developed in
Ref.~\cite{ledoussal02}, shown in Fig.~\ref{fig-mobility}. As is
the case for the model with SR disorder, our model exhibits the
so-called dimension reduction,  both in the statics and in the
dynamics. The naive perturbation theory obtained taking the
functions $\Delta_i(u)$ analytic at $u=0$ leads to the same result
as that computed from the Gaussian theory setting
$\Delta_i(u)=\Delta_i(0)$. In the limit $m\to 0$, the two-point
function then reads to all orders:
\begin{eqnarray} \label{dim-reduction}
 \overline{u_{qt} u_{-q t'}} = \frac{\Delta_1(0)}{c^2q^4}+\frac{\Delta_2(0)}{c^2q^{4+d-a}}.
\end{eqnarray}
The first term in Eq.~(\ref{dim-reduction}) dominates in the limit
$q \to 0$ for $a\ge d $, and  LR disorder is irrelevant in this case, while
the last term dominates for $a<d$. Equation~(\ref{dim-reduction}) results in
$\zeta=(4-d)/2$ for $a\ge d $ and  $\zeta=(4-a)/2$ for $a<d$, which are known to be incorrect.
The physical reason for this is the existence of a large number of metastable  states.
The roughness exponent can be estimated using Flory arguments
setting $u\sim x^{\zeta}$. Then the gradient term scales as  $\nabla^2 u_x \sim x^{\zeta-2}$.
The  pinning force for SR disorder scales as $F(x,u_x)\sim x^{-(d+\zeta)/2}$ and for LR
disorder as $F(x,u_x)\sim x^{-(a+\zeta)/2}$. Therefore, in the regime where the behavior
is governed by SR disorder, the Flory estimate gives for RF disorder the Imry-Ma
value  $\zeta_{\mathrm{SR}}^{\mathrm{F}}=(4-d)/3$ while for LR RF disorder we get
 $\zeta_{\mathrm{LR}}^{\mathrm{F}}=(4-a)/3$. A similar argument
 constructed from the potential correlators $R_i(u)$ yields the
 Flory estimates $\zeta_{\mathrm{SR}}^{\mathrm{F}}=(4-d)/5$ and
 $\zeta_{\mathrm{LR}}^{\mathrm{F}}=(4-a)/5$, respectively, for
 the case of random bond disorder. To obtain
 corrections  to the Flory values, the  FRG developed in
Refs.~\cite{fisher86,nstl92,lnst07,narayan-fisher93,chauve01,ledoussal02} will be employed.
The solution is nontrivial because the renormalized disorder becomes
nonanalytic above the Larkin scale, and one has to deal with a
nonanalytic field theory. Here we generalize this approach to the
case of LR correlated disorder.

\section{Functional Renormalization}
\label{sec5}

We now consider the  renormalization of model (\ref{action}). The subtleties
 arising for the  correlator (\ref{cdw-cumulant}) will  be
discussed briefly at the end. We carry out perturbation theory in the bare
disorder correlators $\Delta_{i0}(u)$ and then introduce the
renormalized correlators $\Delta_{i}(u)$. We will suppress the
subscript "0"  to avoid an overly complicated notation. According to
the standard renormalization program, we compute the effective action
to one-loop order. Here we adopt the dimensional  regularization of
integrals and employ the minimal subtraction scheme to compute the
renormalized quantities and absorb the poles in  $\varepsilon=4-d$
and $\delta=4-a$ into multiplicative \textit{Z} factors. When derivatives of
the $\Delta_i$ at $u=0$ occur, in the dynamics (i.e., at the
depinning transition for dynamical quantities) they are taken at
$u=0^+$ as can be justified exactly for $N=1$. In the statics, the
treatment is more subtle (as discussed in two-loop studies
\cite{ledoussal04}) but is not needed in the present one-loop study.

Let us firstly consider the first-order terms generated by expansion of $e^{-S}$ in
disorder. These terms are given by diagrams
$d$ and $e$ shown in Fig.~\ref{fig-mobility}. We start from
\begin{eqnarray}
&& \!\!\!\!\!\!\!\!\!\! \int_{t>t',x} i \hat{u}_{xt} \Delta_1(u_{xt}-u_{xt'}) i \hat{u}_{xt'}
  \nonumber \\
&&  +   \int_{t>t',x,x'} i \hat{u}_{xt} \Delta_2(u_{xt}-u_{x't'})g(x-x') i \hat{u}_{x't'}.
\end{eqnarray}
Expanding $\Delta_i(u)$ in a Taylor series and contracting one $i\hat{u}$,
 we obtain the leading corrections to the threshold force,
friction and elasticity. The terms giving the threshold force to leading order are
\begin{eqnarray}
&& \!\!\!\!\!\!\!\!\!\! \int_{t>t',x} i \hat{u}_{xt} \Delta_1'(0^+) R_{x=0,t-t'}
\nonumber \\
&&
 +   \int_{t>t',x,x'} i \hat{u}_{xt} \Delta_2'(0^+)g(x-x') R_{x-x',t-t'}. \ \
\end{eqnarray}
They are strongly uv diverging ($\sim \Lambda^{d-2}+\Lambda^{a-2}$), and thus are
nonuniversal.
The terms proportional to $\Delta_i''(0^+)$ can be rewritten as corrections to
friction and elasticity using the expansion
\begin{eqnarray} 
&&  \!\!\!\!\!\!\!\!\!\! u_{xt}-u_{x't'} = (t-t') \partial_t  u_{xt}
+ (x-x')_{i}\frac {\partial}{\partial x_{i}} u_{xt}
\nonumber \\
&&  + (x-x')_i (x-x')_j \frac12 \frac{\partial ^2u_{xt}}{\partial x_i \partial x_j}
  + \mathcal{O}(\Delta t^2, \Delta x^3).
  \label{expansion}
\end{eqnarray}
The first term in Eq.~(\ref{expansion}) gives the correction to friction,
\begin{eqnarray} \label{correction-to-friction}
 \delta \eta  &=& - \Delta_1''(0^+)\int_t t R_{x=0,t}-\Delta_2''(0^+)\int_{xt} t R_{x,t}
 g(x) \nonumber \\
&=& - {\eta}\left[ \hat{\Delta}_1''(0^+) I_{1}+\hat{\Delta}_2''(0^+)
I_{2}\right],
\end{eqnarray}
where we have introduced $\hat{\Delta}_i(u)=\Delta_i(u)/c^2$. The
one-loop integrals $I_1$ and $I_2$ diverge logarithmically and for
$\varepsilon,\delta \to 0$ read
\begin{eqnarray}
  I_1&=&\int_q \frac1{(q^2+\hat{m}^2)^2} = K_4 \frac{\hat{m}^{-\varepsilon}}{\varepsilon}
   + \mathcal{O}(1), \label{I1} \\
  I_2&=&\int_q  \frac{q^{a-d}}{(q^2+\hat{m}^2)^2}=K_4 \frac{\hat{m}^{-\delta}}{\delta}
   + \mathcal{O}(1),\label{I2}
\end{eqnarray}
where we set $\hat{m}=m/\sqrt{c}$ and $K_d$ is the area of a $d$-dimensional sphere divided by $(2\pi)^d$.
To remove the poles in the mobility, we introduce the corresponding $Z$ factor
$\eta_{\mathrm{R}}=Z_{\eta}^{-1}[\Delta_i]\eta$, which to one-loop order is given by
\begin{equation}
  Z_{\eta}^{-1}=1-  \hat{\Delta}_1''(0^+) I_{1}-\hat{\Delta}_2''(0^+) I_{2}.
\end{equation}

\begin{figure}[tbp]
\includegraphics[clip,width=2.2in]{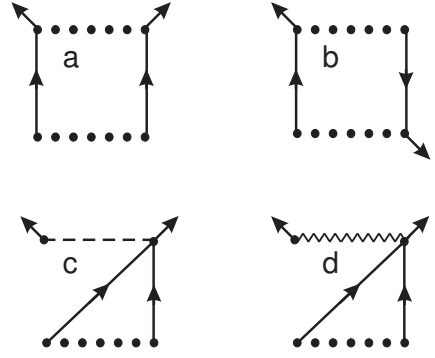}
\caption{One-loop diagrams correcting disorder. The dotted line corresponds to
either SR disorder vertex (dashed line) or to LR disorder vertex (wavy line). Diagrams of type
 $a$, $b$, and $c$ contribute to SR disorder.  Only diagrams of type $d$ correct the
 LR disorder. }
\label{fig-disorder}
\end{figure}

In the absence of  LR correlated disorder, the elasticity remains
uncorrected to all orders due to the STS, while here the correction
reads
\begin{eqnarray}
 \delta c &=&  \frac{1}{2d} \Delta_2''(0)\int_{xt} x^2 R_{x,t} g(x) \nonumber \\
   &=&  -\frac{1}{2d} \Delta_2''(0)\int_{q} g(q)\nabla^2_q \frac1{cq^2+m^2} \nonumber \\
    &=& - c \frac{K_d}{d}  \hat{\Delta}_2''(0)\frac{\varepsilon}{\delta}
    \hat{m}^{-\delta} . \ \ \ \ \
 \label{correction-to-elasticity}
\end{eqnarray}
We have not set the second derivative at $0^+$ as $\Delta_2$ remains
analytic as is discussed below. Furthermore, the correction to
elasticity (\ref{correction-to-elasticity}) is finite for
$\varepsilon, \delta \to 0$, and thus $c$ does not acquire an
anomalous dimension. However, we expect corrections at two-loop order.
If this is the case, one has to introduce a $Z$-factor that
renormalizes elasticity: $c_R=Z_c^{-1}[\Delta_i]c$ with
$Z_c=1+\mathcal{O}(\Delta_i^2)$.

In principle, due to the lack of STS, the KPZ term $\lambda(\nabla
u_{xt})^2 $ breaking the symmetry $u \to -u$ can be  generated in
the equation of motion (\ref{eq-motion}) at the depinning
transition. Indeed, diagram $e$ in Fig.~\ref{fig-mobility}, when
expanding $\Delta(u)$ to second order in $u$, using Eq. (\ref{expansion}),
gives
\begin{eqnarray}
\delta \lambda = \frac{1}{2d} \Delta_2'''(0^+)\int_{xt} x^2 R_{x,t} g(x). \label{KPZ}
\end{eqnarray}
Moreover, the term with cubic symmetry ($M=2$) and terms with
higher-order symmetries ($M>2$)  $\lambda_{M}\sum_{i}(\partial_i u_{xt})^{2M}$  can be
generated by diagram $e$,
\begin{eqnarray}
\delta \lambda_M = \frac{1}{d (2M)!} \Delta_2^{(2M+1)}(0^+)\int_{xt}
 R_{x,t} g(x) \sum_i x_i ^{2M} . \label{irrelevant}
\end{eqnarray}
However, as we will show later, if we start from bare analytic disorder distribution,
the LR disorder remains analytic along the  FRG flow and the corresponding
FP value $\Delta_2^*(u)$
is also an analytic function. Thus terms (\ref{KPZ}) and (\ref{irrelevant})
are zero, provided that they are
absent in the beginning. Moreover, the  terms (\ref{irrelevant}) are irrelevant in the RG sense for $M>2$
[but not the KPZ term (\ref{KPZ}); see Ref.~\cite{ledoussal03-2}].
This proves that our bare model (\ref{action}) is a minimal model for the description
of elastic manifolds in a random media with LR correlated disorder.

The corrections to disorder are given by the diagrams shown in Fig.~\ref{fig-disorder}.
The corresponding expressions read
\begin{eqnarray}
&& \!\!\!\!\!\!\!\!\!\!\!\!\!
  {\delta}^1 \hat{\Delta}_1(u) = -\{ \hat{\Delta}_1'(u)^2
   +[\hat{\Delta}_1(u)-\hat{\Delta}_1(0)]\hat{\Delta}_1''(u) \} I_1   \nonumber \\
&&- \{ 2 \hat{\Delta}_1'(u)\hat{\Delta}_2'(u)^2 +[\hat{\Delta}_2(u)-\hat{\Delta}_2(0)]
    \hat{\Delta}_1''(u)\nonumber \\
&&+ \hat{\Delta}_1(u)\hat{\Delta}_2''(u) \} I_2
- [ \hat{\Delta}_2'(u)^2 +\hat{\Delta}_2(u)\hat{\Delta}_2''(u) ] I_3, \ \ \ \\
&& \!\!\!\!\!\!\!\!\!\!\!\!\!
  {\delta}^1 \hat{\Delta}_2(u) = - \hat{\Delta}_1(0)\hat{\Delta}_2''(u)
  I_1
  - \hat{\Delta}_2(0) \hat \Delta_2''(u) I_2.
\end{eqnarray}
The one-loop integrals $I_{1}$ and $I_{2}$ have been defined in
Eqs.~(\ref{I1}) and (\ref{I2}), whereas $I_3$ is given by
\begin{equation} \label{I3}
  I_3=\int_q  \frac{q^{2(a-d)}}{(q^2+\hat{m}^2)^2} =\frac{K_4
  \hat{m}^{-2\delta+\varepsilon}}{2\delta-\varepsilon} + \mathcal{O}(1).
\end{equation}
Let us define the
renormalized dimensionless disorder $\Delta_{i}^{\mathrm{R}}$ as
\begin{eqnarray}
 {m}^{\varepsilon}\Delta_{1}^{\mathrm{R}}=\hat{\Delta}_1(u)+\delta^1 \hat{\Delta}_1(u), \\
 {m}^{\delta}\Delta_{2}^{\mathrm{R}}=\hat{\Delta}_2(u)+\delta^1 \hat{\Delta}_2(u).
\end{eqnarray}
The $\beta_i$ functions are defined as the derivative of
$\Delta_i^{\mathrm{R}}(u)$ with respect to the mass $m$ at fixed
bare disorder $\Delta_i(u)$. In order to attain a fixed point, it is
necessary to rescale the field $u$ by $m^{\zeta}$ and write the
$\beta$ functions for the functions $\tilde{\Delta}_i=: K_4
m^{-2\zeta}\Delta_i^{\mathrm{R}}(um^{\zeta})$,
\begin{eqnarray}
\partial_{\ell} \tilde{\Delta}_1(u) &=&(\varepsilon-2\zeta) \tilde{\Delta}_1(u) +\zeta u \tilde{\Delta}_1'(u) \nonumber \\
&-&\frac12 \frac{d^2}{d u^2}[\tilde{\Delta}_1(u)+\tilde{\Delta}_2(u)]^2
 + A\tilde{\Delta}_1''(u), \label{frg-del-1} \\
\partial_{\ell} \tilde{\Delta}_2(u) &=& (\delta-2\zeta)\tilde{\Delta}_2(u) +
 \zeta u \tilde{\Delta}_2'(u)
  + A \tilde{\Delta}_2''(u),\qquad
\label{frg-del-2}
\end{eqnarray}
where $A = [\tilde{\Delta}_1(0) +\tilde{\Delta}_2(0)]$ and
$\partial_{\ell}:= - m \frac{\partial}{\partial {m}}$.

The scaling behavior of the system
is controlled by a stable fixed point $[\tilde{\Delta}_1^*(u), \tilde{\Delta}_2^*(u)]$ of
flow equations
(\ref{frg-del-1}) and (\ref{frg-del-2}). To determine the critical exponents,
let us start from power counting following Ref.~\cite{ledoussal03-2}.
The quadratic part of action (\ref{action})
is invariant under $x \to x b $, $t\to t b^z  $, $u \to u  b^{\zeta} $,
$\hat{u} \to \hat{u} b^{2-z-\zeta-d+\psi} $.
Under this transformation, the  mobility, elasticity, and disorder  scale
at the Gaussian FP  as $c \sim b^{\psi}$, $\eta \sim  b^{2-z+\psi}$,
$\tilde{\Delta}_1\sim b^{4-d-2\zeta+2\psi}$, and  $\tilde{\Delta}_2\sim b^{4-a-2\zeta+2\psi}$.
Thus  SR disorder becomes relevant for  $\zeta-\psi<(4-d)/2$ while LR disorder
is naively relevant for  $\zeta-\psi<(4-a)/2$. Note that in the
presence of STS, $\psi=0$, and we recover the conditions obtained
at the end of Sec.~\ref{sec3}.
The actual value of $\zeta$ will be fixed by the disorder correlators at the FP.
The elasticity exponent $\psi$ and the dynamic exponent $z$ read
\begin{eqnarray}
  \psi &=& -{m}\frac{d}{d {m}} \ln Z_{c} (\tilde{\Delta}_i)\Big|_0,  \\
   z&=& 2 + \psi+m\frac{d}{d m} \ln Z_{\eta} (\tilde{\Delta}_i)\Big|_0,
\end{eqnarray}
where subscript ``0" means a derivative at constant bare parameters.
To one-loop order this yields
\begin{eqnarray}
  \psi &=& \mathcal{O}(\varepsilon^2,\varepsilon\delta,\delta^2), \\
   z&=& 2-\tilde{\Delta}_1^{*\prime\prime}(0)-\tilde{\Delta}_2^{*\prime\prime}(0). \label{z-exp}
\end{eqnarray}
The scaling relations then read \cite{ledoussal03-2}
\begin{eqnarray}
&&  \nu=\frac1{2-\zeta+\psi}, \label{nu-exp} \\
&&  \beta=\nu(z-\zeta)= \frac{z-\zeta}{2-\zeta+\psi}. \label{beta-exp}
\end{eqnarray}

At zero velocity, the above calculation can be considered as a dynamical
formulation of the equilibrium problem. However, one has to be careful with mapping
the dynamic FRG equations to the static equations, because as  shown in
Ref.~\cite{ledoussal04} the bare relation $\Delta_i=-R''_i(u)$ breaks down
for the SR case at two-loop order. The standard derivation of the FRG equations in the
statics is based on the renormalization of the replicated Hamiltonian.
We have checked that similar to other systems
with only SR disorder,  the static FRG equations for systems with LR disorder
can be obtained from the dynamic flow equations to one-loop order
using the identification $\tilde{\Delta}_i=-R''_i(u)$. They read
\begin{eqnarray}
\partial_{\ell} R_1(u) &=&(\varepsilon-4\zeta) R_1(u) + \zeta u R'_1(u) \nonumber \\
&&+ \frac12 [R_1''(u)+R_2''(u)]^2 + A R_1''(u), \label{frg-R-1} \\
\partial_{\ell} R_2(u) &=&(\delta-4\zeta)R_2(u) + \zeta u R'_2(u) +   A R_2''(u),\qquad
\label{frg-R-2}
\end{eqnarray}
where $A=-[R_1''(0)+R_2''(0)]$.

In the case of the model with correlator given by Eq.~(\ref{cdw-cumulant}), one has
to distinguish between the  transverse and parallel directions, and therefore introduce
corresponding elastic coefficients $c_{\perp}$ and $c_{\parallel}$.
In the transverse direction, disorder is only $\delta$-correlated and as a result the
transverse elasticity is not corrected and can be set to 1. The power counting
shows that the LR disorder is naively relevant for $\delta_1=4-d_{\perp}-a<0$.
The one-loop integrals are logarithmically divergent and  for $\varepsilon, \delta_1 \to 0$
are given by Eqs.~(\ref{I1}), (\ref{I2}), and (\ref{I3}) with $\delta \to \delta_1$.
Thus the above renormalization can be generalized to model (\ref{cdw-cumulant})
if one formally replaces $\delta \to \delta_1$.

Let us show how a nonanalyticity of the disorder appears in the problem. We start from the bare
analytic correlators with $\tilde{\Delta}_i''(0)<0$.
The flow equation for
$y:=-\tilde{\Delta}_1''(0)-\tilde{\Delta}_2''(0)\equiv R_{1}''''(0)+R_{2}''''(0)>0$ reads
\begin{equation} \label{cusp}
  \partial_{\ell} y = \varepsilon y + 3 y^2 + \gamma({m}),
\end{equation}
where $\gamma({m})=(\varepsilon-\delta)\tilde{\Delta}_2''(0)$.
As we show below, the  function $\tilde \Delta_2(u)$ remains analytic  along the whole
FRG flow and at the fixed point (FP). The solution
of Eq.~(\ref{cusp}) for any function $\gamma({m})$ bounded from below blows
up at some finite scale ${m}^*$,  which can be associated with the inverse
Larkin length. This blowup of $y$ corresponds to the generation of a cusp singularity:
$\tilde{\Delta}_1(u)$ becomes nonanalytic at the origin and acquires for ${m}<{m}^*$
a nonzero $\tilde{\Delta}'_1(0^+)$. The precise estimation of the Larkin scale requires the
solution of the pair of  flow equations for both $\tilde{\Delta}_i(u)$.

Before studying different FPs, let us note an important property that
is valid under all conditions: if $\tilde{\Delta}_i(u)$ ($i=1,2$) is a solution of
Eqs.~(\ref{frg-del-1})
and (\ref{frg-del-2}), then $\kappa^2\tilde{\Delta}_i(u/\kappa)$ is also
a solution. Analogously, if $R_i(u)$ is a solution of Eqs.~(\ref{frg-R-1})
and (\ref{frg-R-2}), then $\kappa^4 R_i(u/\kappa)$ is also a solution.
We can use this property to fix the amplitude of the function in the nonperiodic case,
while for the periodic case the solution is unique as the period is fixed.

\section{ Nonperiodic systems: random bond disorder}
\label{sec6}

\begin{figure}[tbp]
\includegraphics[clip,width=3.2in]{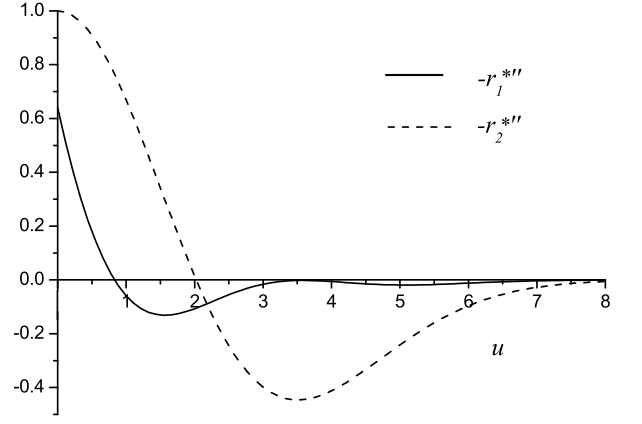}
\caption{Fixed point describing the interface in a medium with long-range correlated
random bond disorder (LR RB FP) for $\hat{\delta}=2$.
The SR part $r_1^*(u)$ is a nonanalytic function with $r_1^{*\prime\prime}(0^+)\neq0$.
The LR part $r_2^{*}(u)$ is an analytic function. Here we report minus their second derivative.}
\label{fig-rbfp}
\end{figure}

In this section, we study the scaling behavior of an
elastic interface in a disordered environment with LR correlated
RB disorder. To this aim, we have to find a stationary solution
(FP) of Eqs.~(\ref{frg-R-1}) and (\ref{frg-R-2}) that decays
exponentially fast at infinity as expected for RB disorder. The SR
RB FP with $R_2(u)=0$, which describes systems with only SR
correlated disorder, was computed numerically in
Refs.~\cite{fisher86,chauve01,ledoussal02}. The corresponding
roughness exponent to one-loop order is given by
$\zeta_{\mathrm{SR}}=0.208\,298\varepsilon+\mathcal{O}(\varepsilon^2)$.
We now look for a LR RB FP with $R_2(u)\neq 0$. Integrating
Eq.~(\ref{frg-R-2}), we obtain
\begin{eqnarray}
\partial_{\ell} \int\limits_0^{\infty} R_2 (u)
&=& (\delta - 5 \zeta)\int\limits_0^{\infty} R_2(u) \label{flow-int-r2}.
\end{eqnarray}
Therefore, the new LR RB FP, if it exists, has
\begin{equation}\label{zeta LRRB}
\zeta_{\mathrm{LRRB}}=
\frac{\delta}5 + \mathcal{O}(\varepsilon^2,\delta^2,\varepsilon\delta).
\end{equation}
The direct inspection of diagrams contributing to
the FRG equation for $R_2$ shows that the higher orders can only be linear in
even derivatives of  $R_2(u)$.  The only term that is linear in $R_2(u)$
comes from the renormalization of the elasticity and can be rewritten
as $2\psi R_2(u)$ to all orders. Therefore, in higher orders we have
\begin{eqnarray}
\partial_{\ell} \int\limits_0^{\infty} R_2 (u)
&=& (\delta - 5 \zeta +2\psi)\int\limits_0^{\infty} R_2(u) \label{flow-int-r2-2},
\end{eqnarray}
and as a consequence, $\int_0^{\infty} du R_2(u)$ is
exactly preserved along the FRG resulting in the exact identity
\begin{equation}\label{zeta LRRB-2}
\zeta_{\mathrm{LRRB}}=\frac{\delta + 2\psi}5.
\end{equation}
Using our freedom to rescale
$R_i(u)$, we introduce $\hat{\delta} :=\delta/\varepsilon$, $R_i(u)
=: \varepsilon r_i(u)$ and  fix $r_1''(0)=:-x$ and $r_2''(0)=-1$,
where $x$ is the parameter to be determined. The stationarity
condition of Eqs.~(\ref{frg-R-1}) and (\ref{frg-R-2}) reads
\begin{eqnarray}
&&\!\!\!\!\!\!\!\!\!\!\!\!\!\!\!
\Big(1-\frac45\hat{\delta}\Big) r_1(u) + \frac{\hat{\delta}}{5} u r'_1(u) \nonumber \\
 && \ \ \ + \frac12 [r_1''(u)+r_2''(u)]^2 + (1+x) r_1''(u) = 0,   \label{frg-r-1} \\
&& \frac{\hat{\delta}}{5} r_2(u) + \frac{\hat{\delta}}{5} u r'_2(u)
+ (1+x) r_2''(u) = 0. \label{frg-r-2}
\end{eqnarray}
Since Eq.~(\ref{frg-r-2}) is linear in $r_2$, it  can be solved for
fixed $x$ by
\begin{eqnarray} \label{r-2}
r_2(u)=\frac{5(1+x)}{\hat{\delta}} \exp\left( -\frac{\hat{\delta}u^2}{10(1+x)} \right).
\end{eqnarray}
From the Taylor expansion of Eq.~(\ref{frg-r-1}) around $u=0$, we
find
\begin{eqnarray}
  r_1(0)&=& \frac{5(1-x^2)}{8\hat{\delta}-10}, \nonumber \\
  r_1'(0)&=& 0, \label{rb-init-cond}
\end{eqnarray}
where the second condition excludes the possibility of a supercusp
(the first line does not diverge since $x=1$ for $\delta=5/4$). Thus
for fixed $\hat{\delta}$ the simultaneous equations (\ref{frg-r-1})
and (\ref{frg-r-2}) have a unique solution for any $x$, but only for
a specific $x$ does the solution $r_1(u)$ decay exponentially fast
to $0$ for large $u$. To determine this value, we employ the shooting
method choosing $x$ as our shooting parameter. For fixed $x$, we
integrate numerically Eq.~(\ref{frg-r-1}) with $r_2(u)$ given by
Eq.~(\ref{r-2}) from $0$ to some large $u_{\mathrm{max}}$ with
initial conditions (\ref{rb-init-cond}). Then the shooting parameter
$x$ can be found by solving numerically the algebraic equation
$r_1(u_{\mathrm{max}};x)=0$. Increasing $u_{\mathrm{max}}$, we
acquire the desired accuracy for $x$ and $r_1(x)$. We were able to
find the numerical solution with reasonable accuracy only for
$\hat{\delta}\ge 1.1$.  The typical FP functions $r_1^*(u)$ and
$r_2^*(u)$ are shown in Fig.~\ref{fig-rbfp}. The actual values of
$x$ obtained by shooting for different $\hat{\delta}$ are summarized
in Table~\ref{tab-stab-rb}.

\begin{table}[tbp]
\caption{Long-range correlated random bond fixed point.
Shooting parameter $x=-r_1''(0)$, the maximal eigenvalue and the universal
amplitude for different values of $\hat{\delta}$.}
\label{tab-stab-rb}%
\begin{ruledtabular}
\begin{tabular}{llll}
$\hat{\delta}$        &    $x(\hat{\delta})$  &  $\lambda_1$                       &$B(\hat{\delta})$ \\ \hline
$1.1$                 &    $1.931986$         &                                    &   $33.89$        \\
$1.2$                 &    $1.121722$         &  $-0.160$                          &   $31.37$        \\
$1.3$                 &    $0.922046$         &  $-0.262$                          &   $31.64$        \\
$1.4$                 &    $0.825747$         &  $-0.365$                          &   $32.41$        \\
$1.5$\footnotemark[1] &    $0.766976$         &  $-0.469$                          &   $33.34$        \\
$2.0$\footnotemark[2] &    $0.639151$         &  $-1$                          &   $38.44$        \\
$3.0$\footnotemark[3] &    $0.562357$         &  $-2.120$                          &   $48.10$        \\
$\infty$              &    $0.463619$         &                                    &   $\infty$       \\ \hline
\end{tabular}
\end{ruledtabular}
\footnotetext[1]{Random lines in a planar interface ($d=2$, $a=1$).
} \footnotetext[2]{Random lines in a 3d manifold ($d=3$, $a=2$).}
\footnotetext[3]{Random planes in a 3d manifold ($d=3$, $a=1$).}
%
\end{table}

Let us now check the stability of SR and LR FPs.  To that end we
linearize the FRG equations about each FP. In the vicinity of a FP, the
linearized flow equations have solutions that are pure power laws
in ${m}$, i.e.,\ scale as ${m}^{-\lambda}$ with a discrete spectrum of
eigenvalues $\lambda$.  A stable fixed point has all eigenvalues
$\lambda < 0$.  Substituting $r_i \to r_i^*(u) + z_i(u)$ into the flow
equations and keeping only terms that are linear in $z_i(u)$, we derive
the linearized flow equations  at the FP $\{r_1^{*}(u),r_2^{*}(u)\}$,
\begin{eqnarray}
\!\! (1&-&4\zeta_1) z_1(u) +
 \zeta_1 u z'_1(u)+ [r_1^{*\prime\prime}(u)+r_2^{*\prime\prime}(u)]
 \nonumber \\
 & \times &
[z_1^{\prime\prime}(u)+z_2^{\prime\prime}(u)] + (1+x) z_1^{\prime\prime}(u) \nonumber \\
 &+& A_0 r_1^{*\prime\prime}(u) = \lambda z_1(u), \ \ \ \label{st-eq-1} \\
(\hat{\delta}&-&4\zeta_1) z_2(u) +  \zeta_1 u z'_2(u) \nonumber \\
& +& (1+x) z_2^{\prime\prime}(u) + A_0
r_2^{*\prime\prime}(u) = \lambda z_2(u), \ \ \label{st-eq-2}
\end{eqnarray}
where we have introduced $\zeta=\varepsilon \zeta_1$,
$A_0=-[z_1^{\prime\prime}(0)+z_2^{\prime\prime}(0)]$, and $\lambda$ is
also measured in units of $\varepsilon$ .  Note that because of the
freedom to rescale $r_i(u)$, we always have the eigenmode $z_i^{(0)}$
with marginal eigenvalue $\lambda_0=0$.  As shown in
Ref.~\cite{ledoussal03} for SR RB FP the corresponding eigenfunction
is given by $z_1^{(0)}=u r_1^{*\prime}(u)-4 r_1^*(u),z_2^{(0)}=0$,
while the next eigenvalue $\lambda_1=-1$ corresponds to
$z_1^{(1)}=\zeta^{\mathrm{SR}}_1 u
r_1^{*\prime}(u)+(1-4\zeta_1^{\mathrm{SR}})r_1^*(u), z_2^{(0)}=0$.
Here $\{r_1^{*},r_2^*=0\}$ is the SR RB FP and the Taylor expansion of
the function $r_1^{*}$ can be found in Ref.~\cite{ledoussal03}.  Thus
the SR RB FP is stable in the SR disorder subspace ($r_2=z_2=0$).  Let
us check its stability with respect to introduction of LR correlated
disorder.  From Eq.~(\ref{st-eq-2}) it follows that the maximal
eigenvalue
$\lambda_{\mathrm{max}}=\hat{\delta}-5\zeta_1^{\mathrm{SR}}$
corresponds to the exponential eigenfunction
$z_2(u)=\exp(-\zeta_1^{\mathrm{SR}} u^2/|2r_1^{*\prime\prime}(0)|)$
with $r_1^{*\prime\prime}(0)=-0.577$ for SR RB FP.  As a consequence,
the LR correlated disorder destabilizes the SR RB FP if
$\hat{\delta}>5\zeta_1^{\mathrm{SR}}\approx 1.041$, or equivalently,
using Eq. (\ref{zeta LRRB}), if
$\zeta^{\mathrm{SR}}<\zeta^{\mathrm{LR}}$. This criterion was of
course expected.

\begin{figure}[tbp]
\includegraphics[clip,width=3.2in]{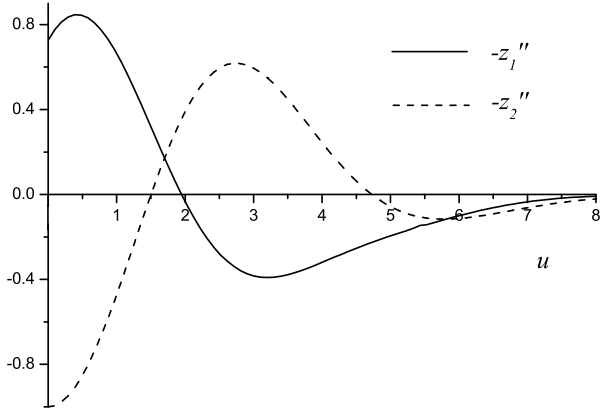}
\caption{Second derivative of eigenfunctions $z_1(u)$ and $z_2(u)$ computed at
 the LR RB FP for $\hat{\delta}=2$.}
\label{fig-rbfpst}
\end{figure}

We now check the stability of the LR RB FP $\{r_1^{*}(u),r_2^{*}(u)\neq0\}$.
It also has a marginal eigenvalue
$\lambda_0=0$ with eigenfunctions given by $z_i^{(0)}= u r_i^{*\prime}(u)-4r_i^*(u)$ that
can be checked by direct substitution into Eqs.~(\ref{st-eq-1}) and (\ref{st-eq-2}).
Equation~(\ref{st-eq-2}) allows for an analytical solution that reads
\begin{eqnarray}
z_2(u)=\frac{5A_0}{2\hat{\delta}+5\lambda}\left[ \frac{\hat{\delta} u^2}{5(1+x)}-1  \right]
\exp\left( - \frac{\hat{\delta}u^2}{10(1+x)} \right). \label{st-rb-y2}
\end{eqnarray}
We are free to fix the length of the eigenvectors, for instance, by the condition
$z_2''(0)=1$, which gives
\begin{eqnarray}
A_0=\frac1{3\hat{\delta}}(1+x)[2\hat{\delta}+5\lambda].
\end{eqnarray}
Thus to find the eigenvalue $\lambda$ and the eigenfunction $z_1$,
we have to solve Eq.~(\ref{st-eq-1}) with condition $z_1''(0)=-1-A_0$ and
require an exponentially fast decay of the solution at large $u$.
The only case for which we succeeded to construct the  solution analytically
is $\hat{\delta}=2$, which is depicted in Fig.~\ref{fig-rbfpst}.
It has eigenvalue  $\lambda=-1$ and reads
\begin{eqnarray}
z_1(u)&=&-\frac13 u r_1^{*\prime}(u)+\frac12 r_1^{*}(u)+\frac56 r_2^{*}(u), \\
z_2(u)&=&-\frac13 [ u r_2^{*\prime}(u)+ r_2^{*}(u)].
\end{eqnarray}
For other values of $\hat{\delta}$ we solve Eq.~(\ref{st-eq-1}) numerically
using  $\lambda$ as a shooting parameter and require an  exponentially fast decay
of  $z_1(u)$ for large $u$.
To compute the numerical solution, we need the initial conditions.
Expanding Eq.~(\ref{st-eq-1}) in a Taylor series, we obtain
\begin{eqnarray}
z_1(0)&= & \frac{5[x^2(2\hat{\delta}+5\lambda) +5x(\hat{\delta}+\lambda) +3\hat{\delta}]
 }{3\hat{\delta}(5-4\hat{\delta}-5\lambda)}, \\
z_1'(0)&=& 0.
\end{eqnarray}
Apart from the   marginal  eigenvalue $\lambda_{0}=0$, the largest eigenvalue is $\lambda_{1}$.
It is shown  for different $\hat{\delta}>1.1$ in Table~\ref{tab-stab-rb}.
The negative sign of $\lambda_1$ reflects the stability of the LR RB FP.
For $\hat{\delta}\le 1.1$ we failed to compute the numerical solution with reasonable
accuracy. However,
the largest eigenvalue computed at LR RB FP $\lambda_1$ tends to $0$
for $\hat{\delta}\to 1.1$ and  the SR RB FP becomes unstable for $\hat{\delta}> 1.041$
with respect to LR-correlated disorder. Thus we
expect that the LR RB FP is stable for $\hat{\delta}> 1.041$.
Moreover, the largest eigenvalue within accessible accuracy is well approximated
by $\lambda_1=0.1917(\zeta_1^{\mathrm{SR}}-\zeta_1^{\mathrm{LR}})$, which
gives $\lambda_1=-0.06$ for  $\hat{\delta}=1.1$.

Besides the roughness exponent, there is another universal quantity that is of interest.
This is the displacement correlation function, which behaves like
\begin{equation} \label{amplitude-def}
  \overline{u_q u_{-q}} = \mathcal{A}_d\, q^{-(d+2\zeta)}.
\end{equation}
Let us show that in contrast to systems with only SR-correlated
disorder,  this system, whose behavior is controlled  by the LR RB
FP, has a universal amplitude $\mathcal{A}_d$. Indeed, according to
Eq. (\ref{flow-int-r2}), the integral $\int du\,
R_2^{\mathrm{tr}}(u)$ is preserved along the flow and is fixed to
its bare value $Q$, where we have introduced the actual renormalized
correlator $R^{\mathrm{tr}}_2$ which is connected to $R_2$ given by
Eq.~(\ref{r-2}) by the relation $R_i^{\mathrm{tr}}=\kappa^4
R_i(u/\kappa)$ with $\kappa$ given by
\begin{equation}\label{Q-par}
  \kappa = \frac{Q^{1/5}}{(2\pi)^{1/10}} \left( \frac{\hat{\delta}}{5(1+x)} \right)^{3/10},
\end{equation}
where we used $\int du R_2^{\mathrm{tr}}(u)=Q$.  Then the amplitude
can be written to one-loop order as follows \cite{ledoussal04}:
\begin{eqnarray}
  \mathcal{A}_d&=&\frac1{K_4}[-R_1^{\mathrm{tr}\prime\prime}(0)-
  R_2^{\mathrm{tr}\prime\prime}(0)] \nonumber \\
   &=&\frac{\kappa^2}{K_4}(1+x)=Q^{2/5}B(\hat{\delta}),
  \label{amplitude}
\end{eqnarray}
where we have introduced the universal function
\begin{equation}\label{amplitude-B}
 B(\hat{\delta}) = \frac{8\pi^2}{(2\pi)^{1/5}}(1+x(\hat{\delta}))^{2/5}
 \left( \frac{\hat{\delta}}{5} \right)^{3/5}.
\end{equation}
Values for $x(\hat \delta)$ and for  $B(\hat{\delta})$ for different $\hat{\delta}$ are shown
in Table~\ref{tab-stab-rb}.

\section{Nonperiodic systems: random field disorder}
\label{sec7}

We now address the problem of an elastic interface in a medium with
LR-correlated RF disorder.  We expect that similar to systems with
uncorrelated disorder, this universality class also describes the
depinning transition. To see that systems with RB disorder flow in the
dynamics to the RF FP, one has to include either effects of a finite
velocity or consider two-loop contributions, which go beyond of the
scope of the present work; but we expect the mechanism to be the same
as in Ref.~\cite{ledoussal02}.

Let us look for a solution of Eqs.~(\ref{frg-del-1}) and
(\ref{frg-del-2}), which decays exponentially fast at infinity as
expected for RF disorder.  From Eq.~(\ref{frg-del-2}) it follows that
(hereafter we drop the tilde on $\Delta_i$)
\begin{eqnarray}
\partial_{\ell}  \int\limits_0^{\infty} {\Delta}_2 (u)
&=& (\delta - 3 \zeta) \int\limits_0^{\infty} {\Delta}_2(u) \label{flow-int-delta2}.
\end{eqnarray}
Therefore, $\int_0^{\infty} {\Delta}_2(u)$ is preserved along the FRG
flow fixing the roughness exponent to
\begin{equation}
  \zeta_{\mathrm{LRRF}}=\frac{\delta}3 + \mathcal{O}(\varepsilon^2,\delta^2,\varepsilon\delta),
\end{equation}
which coincides with the Flory estimate.  Introducing
${\Delta}_i(u)=\varepsilon y_i(u)$, $\zeta=\varepsilon\zeta_1$ and
fixing $y_1(0)=x$, $y_2(0)=1$, we can rewrite the stationary form of
Eqs.~(\ref{frg-del-1}) and (\ref{frg-del-2}) as follows
($\zeta_{1}=\hat \delta/3$):
\begin{eqnarray} 
(1-2\zeta_1)y_1(u) &+& \zeta_1 u y_1'(u)
-\frac12 \frac{d^2}{d u^2}[y_1(u)+y_2(u)]^2 \nonumber \\
& +& [1+x]y_1''(u)=0,
    \label{rf-eq-1} \\
(\hat{\delta}-2\zeta_1)y_2(u) &+& \zeta_1 u y_2'(u) +
  [1+x] y_2''(u)=0. \ \ \ \  \label{rf-eq-2}
\end{eqnarray}
Equation~(\ref{rf-eq-2}) can be solved analytically giving
\begin{eqnarray}
y_2(u)= \exp\left( -\frac{\hat{\delta}u^2}{6(1+x)} \right). \label{y-2}
\end{eqnarray}
Substituting the FP function (\ref{y-2}) in Eq.~(\ref{rf-eq-1}), we obtain a
closed differential equation for the function $y_1(u)$.
Expanding around $u=0$, we find
\begin{eqnarray}
y_1'(0)  &=& -\frac13 \sqrt{9x+3\hat{\delta}-6x\hat{\delta}} , \\
y_1''(0) &=& \frac13 -\frac {\hat{\delta}}{9} \frac{x-2}{x+1}, \label{y1d2}\\
y_{2}'(0)&=& 0,\\
y_2''(0) &=&  -\frac {\hat{\delta}}{3(x+1)} . \label{y2d2}
\end{eqnarray}
Thus we can compute numerically the solution $y_1(u)$ for any fixed
$\hat{\delta}$ and $y_1(0)\equiv x$, however only for special values
of $x$ does this solution decay exponentially at infinity.  The
corresponding solution can be computed using the shooting method as
described above, using $x$ as a shooting parameter (see
Table~\ref{tab-stab-rf}).

\begin{figure}[tbp]
\includegraphics[clip,width=3.2in]{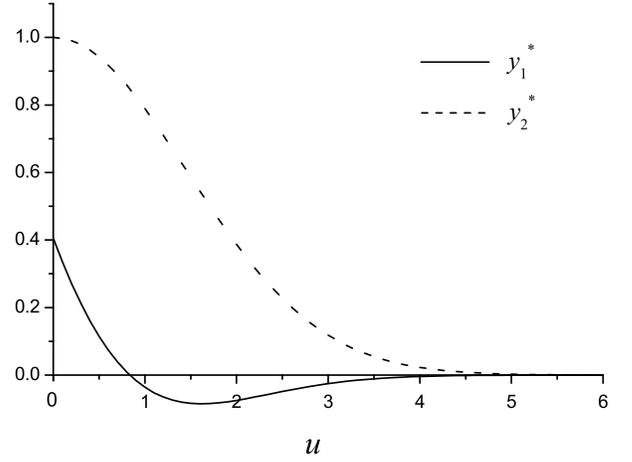}
\caption{ Fixed point describing the interface in a medium with
long-range correlated random field disorder  (LR RF FP) for $\hat{\delta}=2$.
 The SR correlator $y_1^*(u)$ has a cusp at origin
 and formally corresponds to RB type of correlation in direction $u$. The LR correlator
 $y_2^*(u)$ is an analytic function of RF type.} \label{fig-rffp}
\end{figure}

A pair of typical FP functions is shown in Fig. \ref{fig-rffp}.
Surprisingly, the function $y_1(u)$ obtained by shooting satisfies
$\int_0^{\infty}du\ y_1(u)=0$, characteristic for RB-type correlations
along the $u$ direction.  In other words, the LR RF FP is in fact of
mixed type: RB for the SR part and RF for the LR part of the disorder
correlator.  This can be understood as follows: Consider the flow of
$\int_{0}^{\infty}du\, y_{1} (u)$. It is obtained by integrating the
l.h.s.\ of Eq.~(\ref{rf-eq-1}) from 0 to infinity, and by inserting
$\zeta_{1}=\hat{\delta}/3$:
\begin{eqnarray}
&& \!\!\!\!\!\!\!\!\! \partial_{\ell}\int_{0}^{\infty}du\, y_{1} (u) = \left(1-\hat{\delta}
\right) \int_{0}^{\infty}du\, y_{1} (u)  \nonumber \\
&& +\frac{1}{2}
\frac{d}{du}\left[y_{1} (u) +y_{2} (u) \right]^{2}\big|_{u=0} - (1+x) y_{1}' (0)
\label{k222},
\end{eqnarray}
where we have used the fact that most terms in the FRG-equation
(\ref{rf-eq-1}) are total derivatives. Finally, we remark that the second
line of Eq.~(\ref{k222}) cancels exactly provided that $y_2(u)$
is an analytical function, leaving us with
\begin{equation}\label{k1}
\partial_{\ell}\int_{0}^{\infty}du\, y_{1} (u) = \left(1-\hat{\delta}
\right) \int_{0}^{\infty}du\, y_{1} (u) .
\end{equation}
This means that for LR-correlated disorder, i.e.,\ $\hat{\delta}>1$,
the integral of $y_{1}$ indeed scales to 0. A nontrivial fixed
point is possible at two-loop order for depinning. We remind the reader that in
\cite{ledoussal02} it was shown that at two-loop order and for
SR-correlated disorder, new terms arise in the FRG equation, which
do {\em not} integrate to 0. Indeed, this is the mechanism that
leads to a breakdown of the result $\zeta_{\mathrm{SR
RF}}=\varepsilon/3$ at depinning. The same terms will appear here.
We expect that the additional diagrams due to LR correlations do not
exactly cancel these terms, especially since these terms are
proportional to the derivative at the cusp, and LR disorder will
probably remain analytic, thus not contribute to the anomalous
terms. These considerations let us expect that at two-loop order the
integral of $y_{1} (u)$ will be small, but nonzero.

\begin{table}[tbp]
\caption{Long-range correlated random field disorder.
Shooting parameter $x=y_1(0)$, the maximal eigenvalue and the universal
amplitude for different values of $\hat{\delta}$. }
\label{tab-stab-rf}%
\begin{ruledtabular}
\begin{tabular}{llll}
$\hat{\delta}$        &    $x(\hat{\delta})$  &  $\lambda_1$ &$B(\hat{\delta})$                       \\ \hline
$1.1$                 &    $0.562872$         &  $-0.1$      &   $140.43$                              \\
$1.2$                 &    $0.525082$         &  $-0.2$      &   $142.23$                              \\
$1.3$                 &    $0.496948$         &  $-0.3$      &   $144.27$                              \\
$1.4$                 &    $0.475110$         &  $-0.4$      &   $146.44$                              \\
$1.5$\footnotemark[1] &    $0.457638$         &  $-0.5$      &   $148.66$                              \\
$2.0$\footnotemark[2] &    $0.404989$         &  $-1.0$      &   $159.66$                              \\
$3.0$\footnotemark[3] &    $0.362329$         &  $-2.0$      &   $179.04$                              \\ \hline
\end{tabular}
\end{ruledtabular}
\footnotetext[1]{Random lines in planar interface ($d=2$, $a=1$). }
\footnotetext[2]{Random lines in a 3d manifold ($d=3$, $a=2$).}
\footnotetext[3]{Random planes in a 3d manifold ($d=3$, $a=1$).}
%
\end{table}

Let us finally check the stability of the SR RF FP and new LR RF FP.
At the SR RF FP, the roughness is given by
$\zeta_{\mathrm{SRRF}}=\varepsilon/3$, and thus we expect the
crossover from the SR universality to LR at $\delta>\varepsilon$,
which follows from the condition
$\zeta_{\mathrm{SRRF}}=\zeta_{\mathrm{LRRF}}$.  To check the stability
of the FPs, we follow the strategy used for the RB case and linearize
the flow equations about the RF FPs. We obtain
\begin{eqnarray}\label{76}
(1&-&2\zeta_1) z_1(u) + \zeta_1 u z'_1(u) - \frac{d^2}{du^2}\Big\{
[y_1^{*}(u)+y_2^{*}(u)] \nonumber \\
 &\times&[z_1(u)+z_2(u)]\Big\} + (1+x) z_1^{\prime\prime}(u) \nonumber \\
&+& A_0
y_1^{*\prime\prime}(u) = \lambda z_1(u),\ \ \ \ \ \ \ \  \label{st-rf-eq-1} \\
(\hat{\delta}&-&2\zeta_1) z_2(u) + \zeta_1 u z'_2(u)+ (1+x) z_2^{\prime\prime}(u)
\nonumber \\
& +& A_0 y_2^{*\prime\prime}(u) = \lambda z_2(u), \ \ \ \ \label{st-rf-eq-2}
\end{eqnarray}
where we have introduced $A_0=z_1(0)+z_2(0)$.  First, we prove our
conclusion on the stability of SR RF FP with respect to LR-correlated
disorder. To that end, we solve Eq.~(\ref{st-eq-2}) assuming that
$\zeta_1=\zeta_1^{\mathrm{SR}}=1/3$ and $x=y_1^{*\mathrm{SR}}(0)=2/9$.
We obtain that
$z_2=\exp[-\zeta_1^{\mathrm{SR}}u^2/(2y_1^{*\mathrm{SR}})]$ and the
corresponding eigenvalue
$\lambda_{\mathrm{max}}=\hat{\delta}-3\zeta_1^{\mathrm{SR}}$. Therefore,
indeed the SR RF FP becomes unstable with respect to LR disorder for
$\delta>\varepsilon$.

\begin{figure}[tbp]
\includegraphics[clip,width=3.2in]{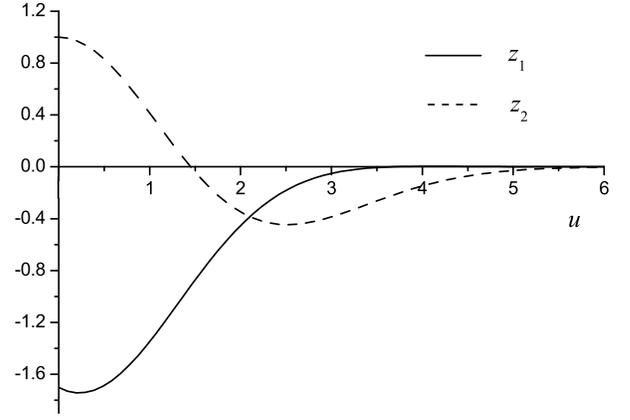}
\caption{Eigenfunctions $z_1(u)$ and $z_2(u)$ computed at the LR RF FP
for $\hat{\delta}=2$.}
\label{fig-rffpst}
\end{figure}

We now focus on the stability of the LR RF FP.  Analysis of the
linearized flow equations (\ref{st-rf-eq-1}) and (\ref{st-rf-eq-2}) shows
that there is at least one eigenvector
$z_i^{(0)}=u\Delta_i^*(u)-2\Delta_i^*$ with marginal eigenvalue
$\lambda_0=0$, which corresponds to the freedom of rescaling.  For
arbitrary $\lambda$, Eq.~(\ref{st-rf-eq-2}) can be solved
analytically,
\begin{eqnarray}
z_2(u)= \frac{A_0\left[\delta{u}^{2}-3(1+x)\right]}{3 (2\delta+3\lambda)(1+x)^2}
\exp\left( -\frac{\hat{\delta}u^2}{6(1+x)} \right). \label{st-rf-y2}
\end{eqnarray}
We are free to fix the length of the eigenvectors, for instance by
the condition $z_2(0)=1$, which gives
\begin{eqnarray}
A_0=-\frac1{\hat{\delta}}(2\hat \delta+3\lambda)(1+x).
\end{eqnarray}
Thus to find the eigenvalue $\lambda$ and the eigenfunction $z_1$, we
have to solve Eq.~(\ref{st-rf-eq-1}) with condition $z_1(0)=A_0-1$ and
require an exponentially fast decay for large $u$.  We need the
initial conditions that can be found by expanding
Eq.~(\ref{st-rf-eq-1}) in a Taylor series,
\begin{eqnarray}
z_1(0)&= & -1 -\frac1{\hat{\delta}}(2\hat \delta+3\lambda)(1+x), \\
z_1'(0)&=& - \left\{ [9\lambda^2+\lambda(12\hat \delta-9)](x+1) +\hat \delta^2(4x+7)
\right. \nonumber \\
&&- \left. \hat \delta(6x+9)\right\} / 2 \hat \delta \sqrt {9x-6x\hat \delta +3\hat \delta}.
\end{eqnarray}
$\hat{\delta}=2$ is the only case in which we succeeded to find a
completely analytical solution (see Fig.~\ref{fig-rffpst}).
It has $\lambda_1=-1$ and reads
\begin{eqnarray}\label{z1}
z_1(u)&=& u y_1^{*\prime}(u)-\frac12 y_1^{*}(u)-\frac32 y_2^{*}(u), \\
z_2(u)&=& u y_2^{*\prime}(u)+ y_2^{*}(u). \label{z2}
\end{eqnarray}
For other values of $\hat{\delta}$, Eq.~(\ref{z2}) remains correct,
while to obtain $z_{1}(u)$, we solve Eq.~(\ref{st-rf-eq-1})
numerically, using $\lambda$ as a shooting parameter.  As can be
seen from Table~\ref{tab-stab-rf}, the largest eigenvalue satisfies
$\lambda_1(\hat{\delta})=1-\hat{\delta}\equiv
3(\zeta_1^{\mathrm{SR}}-\zeta_1^{\mathrm{LR}})$. This result can be
obtained analytically as follows. Integrating Eq.~(\ref{76}) from 0
to $\infty$, we get the condition
\begin{equation}
\int_{0}^{\infty}du\, z_{1}(u) \left( 1- \hat \delta -\lambda\right)
=0,
\end{equation}
proving that as long as $\int z_{1} (u)\neq 0$, one has $\lambda =
1-\hat{\delta } $. Therefore, the LR RF FP is stable for $\delta>\varepsilon$.
Inserting this value of $\lambda$ back into Eq.
(\ref{76}), we obtain after some simplifications
\begin{eqnarray}\label{k2}
0 &=&\frac{\hat \delta}{3}\frac d{du} \left[ u z_{1}(u)\right] - \frac {d^{2}}{du^{2}}  \left[
Y(u) z_{1}(u) + W(u) \right],\nonumber \\
 Y (u)&:=&  y_{1} (u)-y_{1} (0)+y_{2} (u)-y_{2} (0),\\
 W (u)& :=& z_{2} (u) [y_{1} (u)+y_{2} (u)] - y_1(0)[z_1(0)+z_2(0)] . \nonumber
\end{eqnarray}
The first equation can be integrated with the result
\begin{equation}\label{k3}
\frac{\hat \delta}{3}  u z_{1}(u) = \frac{d}{du}  \left[
Y(u) z_{1}(u) + W(u)  \right].
\end{equation}
This is equivalent to
\begin{equation}\label{k4-2}
z_{1} \left[Y'-\frac{\hat{\delta}}{3}u \right] + z_{1}' Y =  - W'.
\end{equation}
The homogenous equation reads
\begin{equation}\label{k5}
\left[\ln z_{1} Y \right]' =\frac{\hat{\delta}}{3} \frac{u }{Y}.
\end{equation}
Its solution is
\begin{equation}\label{k6}
z_{1} (u) = \frac{C }{Y (u)} \exp \left[\frac{\hat{\delta}}{3} \int^{u}_0 ds\, \frac{s
}{Y (s)} \right]
\end{equation}
with some constant $C$. A solution to the inhomogeneous equation is
obtained by replacing $C$ by $C (u)$, and inserting the latter into
Eq.~(\ref{k4-2}). This yields
\begin{equation}
C (u) = - \int^{u}_0 dt  W'(t)  \exp
\left[-\frac{\hat{\delta}}{3} \int^{t}ds\, \frac{s}{Y (s)} \right] .
\end{equation}
Putting together everything, we obtain
\begin{equation}
z_{1} (u) =- \frac{1}{Y
(u)} \int^{u}_{0} dt W '(t) \exp
\left[-\frac{\hat{\delta}}{3} \int^{t}_{u}ds\, \frac{s}{Y (s)} \right].
\end{equation}

Let us now compute the universal amplitude defined by
Eq.~(\ref{amplitude-def}). According to Eq.~(\ref{flow-int-delta2}),
the integral $\int du\, \Delta_2(u)$ is preserved along the FRG flow
and can be fixed by its bare value $Q$. The relation between the
actual renormalized  disorder
$\Delta_2^{\mathrm{tr}}$ and the rescaled disorder $\Delta_2(u)$
given by Eq.~(\ref{y-2}) reads $\Delta_i^{\mathrm{tr}}=\kappa^2
\Delta_i(u/\kappa)$.  We obtain
\begin{equation}\label{Q-delta}
  \kappa = \frac{Q^{1/3}}{\varepsilon^{1/3}} \left( \frac{2\hat{\delta}}{3\pi(1+x)}
   \right)^{1/6},
\end{equation}
where we have fixed $\int du \Delta_2^{\mathrm{tr}}(u)=Q$.
Then  the amplitude can be written as
\begin{eqnarray}
  \mathcal{A}_d&=&\frac1{K_4}[\Delta_1^{\mathrm{tr}}(0)+
  \Delta_2^{\mathrm{tr}}(0)] \nonumber \\
  &=&\frac{\varepsilon \kappa^2}{K_4}(1+x)=\varepsilon^{1/3}
   Q^{2/3}B(\hat{\delta}),\label{RF-amplitude}
\end{eqnarray}
with the universal functions given by
\begin{equation}\label{RF-amplitude-B}
 B(\hat{\delta}) = {8\pi^2}(1+x)^{2/3} \left( \frac{2\hat{\delta}}{3\pi} \right)^{1/3}.
\end{equation}
Values of $B(\hat{\delta})$ for different $\hat{\delta}$ are shown in
Table~\ref{tab-stab-rf}.

 \textit{Depinning.} We are now in the position to study the depinning transition,
which we expect is controlled by the LR RF FP.
The dynamic critical exponent $z$ defined by Eq.~(\ref{z-exp}) is given to one-loop order by
\begin{equation}
  z = 2-\frac{\varepsilon}3  + \frac{\delta}9 + \mathcal{O}(\varepsilon^2,\delta^2,
   \varepsilon\delta),
\end{equation}
where we have used Eqs.~(\ref{y1d2}) and (\ref{y2d2}), which give
$y_1''(0)+y_2''(0)=1/3-\hat{\delta}/9$.
Other exponents can be computed using scaling relations (\ref{beta-exp})
and (\ref{nu-exp}), for example
\begin{eqnarray}
  \beta = 1-\frac{\varepsilon}6+ \frac{\delta}{18}
  + \mathcal{O}(\varepsilon^2,\delta^2,    \varepsilon\delta).
\end{eqnarray}
It is remarkable that for $\delta > 3\varepsilon$, the exponent
$\beta$ is larger than $1$, and $z$ is larger than $2$. This seems to
imply some different physics - yet to be understood - in the
avalanche process, which makes the motion slower near depinning than
in the SR case. The analyticity of $\Delta_2$ seems to suggest some
smoother motion at large scale, while short-scale motion remains
jerky and avalanche-like. Finally, note that at the SR RF FP,
$z_{\mathrm{SR}}=2-2\varepsilon/9$,
$\beta_{\mathrm{SR}}=1-\varepsilon/9$, and thus, the exponents are
continuous functions of $\varepsilon$ and $\delta$.

\section{Periodic systems}
\label{sec8}

\begin{table}[tbp]
\caption{Periodic systems with LR correlated disorder.
The shooting parameter: $y_1(0)$ and two first eigenvalues
 for different  $\hat{\varepsilon}$.}
\label{tab-periodic}%
\begin{ruledtabular}
\begin{tabular}{llll}
$\hat{\varepsilon}$    &    $y_1(0)$   & $\lambda_1$  & $\lambda_2$  \\  \hline
$0$                    &    $0.00971$  &   $0.0$      &               \\
$1/3$                  &    $0.01089$  &   $0.333$    &  $-4.089$    \\
$1/2$                  &    $0.01183$  &   $0.500$    &  $-0.500$    \\
$2/3$\footnotemark[1]  &    $0.01348$  &   $0.667$    &  $-0.280$    \\
$0.8$                  &    $0.01645$  &   $0.800$    &  $-0.125$    \\
$0.9$                  &    $0.02346$  &   $0.900$    &  $-0.013$    \\
\end{tabular}
\end{ruledtabular}
\footnotetext[1]{Corresponds to $d=2$, $a=1$, i.e.,\ line defects
(e.g., dislocations) along the plane of a CDW.}
\end{table}

\begin{figure}[tbp]
\includegraphics[clip,width=3.6in]{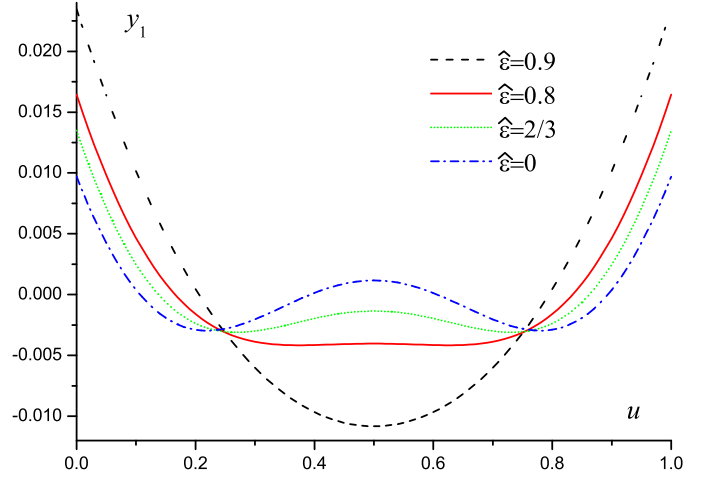}
\caption{(Color online) Fixed point of a periodic system with LR correlated disorder.
 The SR disorder correlator $y_1(u)$ computed for different values of $\hat{\varepsilon}$.}
\label{fig3}
\end{figure}

We now study periodic systems with disorder correlator given by
Eq.~(\ref{model}), which we can refer to as an \textit{XY} model with LR-correlated
defects. The results for CDWs with LR-correlated disorder defined  by
correlator (\ref{cdw-cumulant}) can  then be obtained by substituting
$\delta \to \delta_1=4-d_{\perp}-a$. It is sufficient to consider the system with the
period fixed to 1, since other systems can be related to the latter using the
freedom to rescale.
As a consequence, the roughness exponent for periodic systems is $\zeta=0$.
At variance with interfaces, we introduce reduced parameters
according to $\Delta_i(u)=\delta y_i(u)$, $A=y_1(0)+y_2(0)$, and
$\hat{\varepsilon}=\varepsilon/\delta$. Then the fixed-point equations can be written
as follows:
\begin{eqnarray}
\hat{\varepsilon} y_1(u)
&-& \frac12 \frac{d^2}{d u^2}[y_1(u)+y_2(u)]^2 +  A y_1''(u)=0, \qquad    \label{period-eq-1} \\
 y_2(u) &+&   A y_2''(u)=0. \ \ \ \  \label{period-eq-2}
\end{eqnarray}
Equation~(\ref{period-eq-2}) can be solved analytically. Its solution is
\begin{equation}\label{periodic-y2}
  y_2=y_2(0)\cos(2\pi u), \ \ \ A = 1/(2\pi)^2.
\end{equation}
Equation~(\ref{period-eq-1}) can be solved analytically for
$\hat{\varepsilon}=0$,
\begin{eqnarray}
  y_1&=&y_1(0)+ y_2(0)\left[1-\cos(2\pi u)\right]  \nonumber \\
 &&- \frac1{2\pi}\sqrt{2 y_2(0)\left[1-\cos(2\pi u)\right]}. \label{periodic-y1-eps0}
\end{eqnarray}
The coefficients $y_i(0)$ are determined by potentiality of the $\Delta_i$, i.e.,\ from
the conditions
\begin{equation}\label{periodic-potentislity}
  \int\limits_0^1 du\ y_1(u)=\int\limits_0^1 du\ y_2(u)=0,
\end{equation}
and the identity  $y_1(0)+y_2(0)=1/(2\pi)^2$. They read
\begin{eqnarray}
y_1(0)&=&1/(2\pi)^2 -1/64, \\
y_2(0)&=&1/64.
\end{eqnarray}
For $\hat{\varepsilon}>0$, Eq.~(\ref{period-eq-1}) can be written in the following form:
\begin{equation}\label{periodic-numeric-eq}
\hat{\varepsilon} y_1(u)
- \frac12 \frac{d^2}{d u^2}\left\{\left[y_1(u)+y_2(u)\right]^2 -
 \frac{y_1(u)}{\pi^2} \right\}=0,
\end{equation}
where  $y_2(u)$ is given by Eq.~(\ref{periodic-y2}) with $y_2(0)=1/(2\pi)^2 - y_1(0)$.
Expanding  Eq.~(\ref{periodic-numeric-eq}) in a Taylor series about $u=0$,
we find that $y_1'(0)=-\sqrt{1/(2\pi)^2-y_1(0)(1-\hat{\varepsilon})}$. Thus for
any fixed $0\le\hat{\varepsilon}<1$ and  $y_1(0)$ we have only one solution $y_1(u)$,
but only for a specific $y_1(0)$ this solution fulfills the condition
$y_1(1)=y_1(0)$. To find this value, we employ the shooting method using
$y_1(0)$ as a shooting parameter. The values of $y_1(0)$ computed for
 different $\hat{\varepsilon}$ are summarized in Table~\ref{tab-periodic}.  The
corresponding eigenfunctions $y_1(u)$ are depicted in Fig.~\ref{fig3}.

While the roughness exponent is zero, the system forms a Bragg glass phase
with a slow growth of the displacements according to
\begin{equation}\label{periodic displacement}
  \overline{(u_x-u_0)^2}= \mathcal{A}_d \ln |x|,
\end{equation}
where $\mathcal{A}_d$ is a  universal amplitude, which to one-loop
order is given by
\begin{equation}\label{periodic amplitude}
  \mathcal{A}_d^{(\mathrm{LR})}  =
  \frac{2K_d}{K_4}[\Delta_1^*(0)+\Delta_2^*(0)] = \frac{\delta}{2\pi^2},
\end{equation}
where we have restored the factor of $1/K_4$ previously absorbed in $\Delta_i$.
The SR periodic FP is characterized by
$\mathcal{A}_d^{(\mathrm{SR})}=\varepsilon/18+\mathcal{O}(\varepsilon^2)$. It is interesting
to compare Eq. (\ref{periodic amplitude}) with the prediction of the Gaussian
variational approximation
for the SR disorder case $\mathcal{A}_{d,\mathrm{GVA}}^{(\mathrm{SR})}=\varepsilon/(2\pi^2)$.
We expect the crossover between LR and SR FPs at
$\mathcal{A}_d^{(\mathrm{LR})}=\mathcal{A}_d^{(\mathrm{SR})}$,
i.e.,\ LR disorder to be relevant for
\begin{equation} \label{period-st-arg}
\frac{\varepsilon}\delta<\frac{9}{\pi^2}\approx 0.912.
\end{equation}

\begin{figure}[tbp]
\includegraphics[clip,width=3.3in]{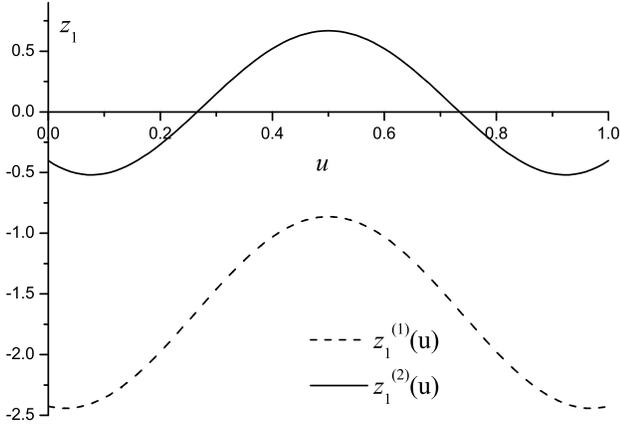}
\caption{Two first eigenvectors computed at the LR periodic  FP.
 Eigenfunctions  $z_1^{(1)}(u)$ and $z_1^{(2)}(u)$ for  $\hat{\varepsilon}=2/3$.  }
\label{period-stab}
\end{figure}

We now check the stability of the SR and LR periodic FPs.
The flow equations linearized  about the FP   read
\begin{eqnarray}
 \hat{\varepsilon} z_1(u)  &-& \frac{d^2}{du^2}
\Big\{ [y_1^{*}(u)+y_2^{*}(u)][z_1(u)+z_2(u)]\Big\} \nonumber \\
 & + & A z_1^{\prime\prime}(u) + A_0 y_1^{*\prime\prime}(u)
   = \lambda z_1(u),   \label{st-period-eq1} \\
 z_2(u) &+&  A z_2^{\prime\prime}(u) + A_0 y_2^{*\prime\prime}(u) = \lambda z_2(u),
  \label{st-period-eq2}
\end{eqnarray}
where  we have introduced $A_0=z_1(0)+z_2(0)$.
Let us recall  that the SR FP is unstable with respect to  nonpotential perturbations even in the subspace  of SR disorder. Indeed, the SR periodic FP,
\begin{equation}\label{k7}
  \Delta_1^*(u)= \frac{\varepsilon}{6}\left[\frac16-u(1-u)\right],\ \ \
  \Delta_2^*(u)=0,
\end{equation}
has in the SR subspace the positive eigenvalue $\lambda_1=1$, corresponding to the nonpotential
eigenfunction $z_1=1$. All other eigenfunctions are potential, i.e.,\ fulfill condition
(\ref{periodic-potentislity}),  and have negative eigenvalues \cite{ledoussal03}.
If we add LR-correlated disorder,
the solution of Eq.~(\ref{st-period-eq2}) yields
\begin{equation}
  z_2(u)=\cos2\pi u. \label{cos}
\end{equation}
The corresponding eigenvalue
$\lambda_{\mathrm{SR}}=1-\hat{\varepsilon}\pi^2/9$ confirms our estimation for the stability
of the SR periodic FP  (\ref{period-st-arg}).
For the LR periodic FP, we still have Eq.~(\ref{cos}) with
\begin{equation}
  A_0=-\frac{\lambda}{1-4\pi^2 y_1(0)}, \ \ \ z_1(0)=A_0-1.
\end{equation}
Equation~(\ref{st-period-eq1}) has a periodic solution only for a discrete set of eigenvalues
$\lambda_i$ (the first two are shown in Table \ref{tab-periodic}). It follows from the table
that $\lambda_1=\hat{\varepsilon}>0$.
In analogy with the SR periodic FP,
the LR periodic FP is unstable with respect to a nonpotential perturbation
corresponding to $\lambda_1$.  The latter is obtained by integrating Eq. (\ref{st-period-eq2})
over one period,
\begin{equation}
(\hat\varepsilon-\lambda) \int_{0}^{1} du\, z_{1}(u) =0.
\end{equation}
As long as the integral does not vanish, this gives the reported eigenvalue
$\lambda_{1}=\hat\varepsilon$.  Indeed, as can be seen  from
Fig.~\ref{period-stab}, we have $\int_0^1 du z_1^{(1)}(u)\neq 0$ and
$\int_0^1 du z_1^{(n)}(u)\neq 0$ for $n\ge 2$.

\textit{Depinning}.
We now focus on the depinning transition of the periodic system with LR-correlated
disorder. At the LR periodic FP, we have
\begin{eqnarray}
y_1''(0) &=& 1+ \frac{\hat{\varepsilon}}3 - 4\pi^2 y_1(0), \\
y_2''(0) &=& - 4\pi^2 y_2(0),
\end{eqnarray}
and thus $y_1''(0)+y_2''(0)={\hat{\varepsilon}}/3$.
The dynamic critical exponent $z$ defined by Eq.~(\ref{z-exp}) reads to one-loop order
\begin{equation}
  z^{\mathrm{LR}} = 2-\frac{\varepsilon}3  + \mathcal{O}(\varepsilon^2,\delta^2,
   \varepsilon\delta).
\end{equation}
Therefore, for periodic systems $z^{\mathrm{LR}}=z^{\mathrm{SR}}$ to one-loop order.

\section{Fully isotropic extended defects}
\label{sec:extended}

\begin{figure}[tbp]
\includegraphics[clip,width=2.2 in]{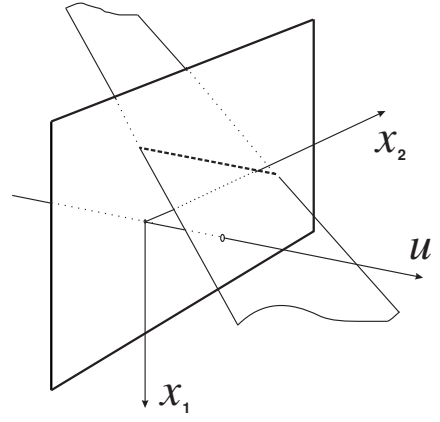}
\caption{2d domain wall moving in 3D magnet with fully isotropic
planar defects. }
\label{fig-dw}
\end{figure}

In this section, we briefly examine the effect of a defect
distribution isotropic in the whole $(x,u)$ space. Consider first
 interfaces in random bond type disorder. From Eqs.
(\ref{disorder potential}) and (\ref{rb}), one finds
\begin{eqnarray}
 R(x,u)= \overline{V_{\mathrm{RB}}(x,u) V_{\mathrm{RB}}(0,0)} \sim
  \frac{v_{\mathrm{LR}}^2}{|x^2+u^2|^{a/2}}, \label{bare-correlator-2}
\end{eqnarray}
and thus the $u$ and $x$ dependences are coupled in the bare
correlator. For the present discussion, we consider $N$, the number
of components of $u$, arbitrary, hence $D=N+d$. We recall that
$a=D-\varepsilon_d$. The question of to which universality class
this model belongs is subtle. It turns out that it does not
correspond to LR disorder in internal space, but rather SR disorder
in internal space and LR disorder in the $u$ direction, hence $R_2=0$
but $R_1(u)$ long range in $u$. To see this, let us consider at fixed
$u$ the integral $\int d^d x R(x,u)$. We can distinguish two
cases, as follows.

\begin{figure}[tbp]
\includegraphics[clip,width=3.2 in]{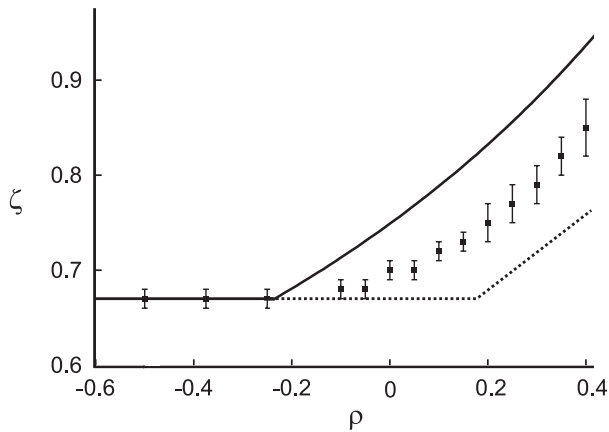}
\caption{ The roughness exponent of the
optimal paths on the plane with isotropically correlated disorder
(data taken from Ref.~\cite{rieger03}). Solid line
$\zeta=3/(4-2\rho)>\zeta_{\mathrm{SR}}=2/3$
is the roughness  exponent of  the
elastic string on the  plane with fully isotropic long-range correlated
disorder. Dashed line $\zeta=(3+2\rho)/5>\zeta_{\mathrm{SR}}$ is
the roughness exponent of the elastic
string on the plane with disorder correlated only along the string. }
\label{heiko}
\end{figure}

(i) For $a>d$, this integral is convergent at large $x$, hence we
clearly have SR disorder in the $x$ direction, and  $R_1(u) \sim |u|^{d-a}$
at large $u$. This, however, is LR disorder in $u$. This case has been
studied using FRG and yields, for $a<a_c(d,N)$, a roughness exponent
given by the Flory value $\zeta(a,d)=(4-d)/(4+a-d)$. The
value $a_c(d,N)$ can be estimated using the value for the SR
disorder roughness exponent, by requiring
$\zeta(a_c(d,N),d)=\zeta_{\mathrm{SRRB}}(d,N)$ (small deviations can arise
as discussed in \cite{vortex}).

(ii) For $a<d$, the situation is more subtle and one may be tempted
to argue, since $\int d^d x R(x,u)$ diverges in the infrared, that
disorder LR in $x$ is produced. This is, however, not the case, as
can be seen on the Fourier transform $R(q,P)$, where $P$ is the
momentum associated with $u$, and $q$ with $x$. One has $R(q,P) \sim
(q^2+P^2)^{(a-d-N)/2}$, which has a well defined limit $R(q=0,P) =
P^{a-d-N}$. This corresponds again, as we argue, to a SR correlator in space
with $R_1(u)-R_1(0) \sim |u|^{d-a}$. As is often the case, LR
models require some trivial subtractions. The subtracted correlator
$R(x,u)-R(x,0)$ has indeed a convergent integral $\sim |u|^{d-a}$ at
large $u$, while subtracting a $u$-independent piece does not
change the model. The critical case $a=d$ is described by the
logarithmic model $R(x,u)-R(x,0) \sim \ln |u|$, which has
$\zeta=(4-d)/4$ in all dimensions \cite{ledou}.

To summarize, isotropic distributions of defects isotropic in the
$(x,u)$ space also yield LR models, but not of the type
(\ref{model}) studied here. For isotropic line defects, one finds
$\zeta=(4-d)/(3+N)$ (i.e., $\zeta=3/4$ for a directed polymer in
$D=1+1$, $\zeta=3/5$ in $D=1+2$, and for an interface $D=2+1$,
$\zeta=2/5$). Isotropic planar defects yield $\zeta=(4-d)/(2+N)$,
hence $\zeta=2/3$ for a $(D=2+1)$-dimensional interface. This case is illustrated
in Fig.~\ref{fig-dw}.
Note that in that case there are infinitely many lines of defects
inside the interface with random directions (the intersections of
the planar defects with the interface gives lines),
but that this does not suffice to create power-law correlations in
internal space, as can be seen from the example in which the planar
defects are orthogonal to the interface.

In Ref.~{\cite{rieger03}}, the universal properties of the optimal paths
on the plane with isotropically correlated random potential which
correlation decays as $r^{2\rho-1}$ were studied using numerical simulations.
This model is believed to belong to the same universality class
as the one-dimensional ($d=1$) elastic interface in a medium
with fully isotropic long-range correlated disorder with $a=1-2\rho$.
The roughness exponent of the shortest paths
computed in Ref.~{\cite{rieger03}} for different values of $\rho$ is
shown in Fig.~\ref{heiko}. In the optimal path model, the elasticity is
generated by disorder and the long-range correlated disorder can generate
the long-range elasticity, which would decrease the path roughness and
account for the deviation from our prediction. The effective elasticity
is expected to be described in terms of the exponent $\psi$,
which computation requires the two-loop consideration.

Finally in the periodic case, such as for CDWs, isotropic disorder in
the full space $(x_\parallel,x_\perp)$ again leads to correlations
(\ref{cdw-cumulant}), but now the function
$g(x_{\parallel}-x_{\parallel}')$ decays exponentially beyond a
length scale set by the disorder period (as can be seen in Fourier
space considering the discrete $P$ modes). Hence the problem is
described by the standard (SR) random periodic class.

\section{Conclusion}
\label{conclusion}

We have studied elastic interfaces and periodic systems  in  a
medium with LR correlated disorder, both  in equilibrium and at
the depinning transition. This type of long-range correlation
exists in the internal space of the manifold, and we have
discussed how it can be realized in terms of extended defects, or
anisotropic defects with a broad distribution of lengths. Using a
dynamic formalism, we derived the FRG flow equations for the SR and
LR parts of the disorder correlator and found three new FPs, which
describe three new universality classes. All new FPs are
characterized by a nonanalytic SR part of the disorder correlator
and an analytic LR part. We have computed the corresponding
exponents and universal amplitudes in a double expansion in
$\epsilon=4-d$ and $\delta=4-a$. For RB type of disorder, we find
that the LR correlation of disorder is relevant for
$\delta>1.041\varepsilon$ and results in the roughness exponent
$\zeta=\delta/5$, while for
$\delta<1.041\varepsilon$ the scaling behavior is controlled by
the SR RB FP with $\zeta=0.208\,298\varepsilon$. We find that the
presence of RF disorder results in a mixed FP with the SR
correlator corresponding formally to RB type of disorder and an
analytic RF LR correlator. The LR RF FP, which is also expected to
control the depinning transition, is stable for
$\delta>\varepsilon$ giving $\zeta=\delta/3$ and
$\beta=1-\varepsilon/6+\delta/18$. The LR correlated periodic FP
is stable for $\varepsilon<0.912\,\delta$ and gives a slow
logarithmic growth of displacements with universal amplitude
$\mathcal{A}_d^{\mathrm{LR}}=\delta^2/(2\pi^2) $. It is remarkable
that this type of disorder yields an exponent $\beta$ for the
velocity-force characteristics  that can be larger than unity and
a dynamical exponent larger than $2$. This striking behavior might
be relevant for experiments, and gives a strong motivation for
numerical studies of the problem, e.g., to understand the nature
of motion at the depinning transition in these systems.

\begin{acknowledgments}
We thank Dima Feldman and Heiko Rieger for useful discussions, and
the Kavli Institute for Theoretical Physics
at the University of California, Santa Barbara  for
hospitality  while this  work  was being finished. This research was
supported in part by the National Science Foundation under Grant
No.~PHY99-07949. A.A.F. acknowledges support from the European Commission
under Contract No.~MIF1-CT-2005-021897. P.L.D. and K.J.W. are
supported by Agence Nationale de la Recherche under Contract
No.~05-BLAN-0099-01.
\end{acknowledgments}

\end{document}